# Quantum–Hydrodynamic Framework of Coherent Terahertz Emission in MXenes Driven via Engineered Femtosecond Waveforms

Ali Asghar Molavi Choobini[1*] Abbas Chimeh[1,2], Jinhui Zhong[3]
[1]Quantum Matter Lab, Department of Physics, College of Science, University of Tehran, Tehran 14399-55961, Iran,
[2]Nexus for Quantum Coherence and Entanglement in Light-Matter Systems (Qcelms), University of Tehran, P.O. Box 14395-547, Tehran, Iran,
[3]Department of Materials Science and Engineering, Southern University of Science and Technology, Shenzhen 518055, China

**Abstract:** Coherent terahertz (THz) emitters based on low-dimensional quantum materials have attracted significant interest for compact broadband photonic technologies. However, theoretical descriptions that consistently connect microscopic quantum dynamics to emitted THz radiation remain limited for MXenes, whose electronic properties are governed by transition-metal d-orbitals and chemically tunable surface terminations. A multiscale density-matrix formalism for coherent THz generation in MXenes driven by engineered three-color femtosecond waveforms is developed. MXene-dependent effective electronic structures are described by parameterized multiband tight-binding Hamiltonians, while ultrafast carrier dynamics are treated using a non-secular Redfield master equation that explicitly retains interband coherence and carrier populations together with effective environment-induced relaxation and dephasing. The transient photocurrent is explicitly decomposed into population- and coherence-driven contributions and linked to the emitted THz field through its time derivative, establishing a direct connection between microscopic quantum dynamics and macroscopic radiation. The results show that environmental coupling does not merely attenuate THz emission, but redistributes coherent and population currents and reshapes the emitted waveform in both the time and frequency domains. Systematic investigations of optical waveform parameters, effective electronic structure, and open-system relaxation identify physically grounded strategies for enhancing broadband THz generation. The proposed methodology establishes a portable microscopic design platform for connecting effective electronic structure, quantum coherence, open-system relaxation, nonlinear carrier transport, and coherent THz emission across different MXene systems and related low-dimensional quantum materials.

*Corresponding author: E-mail address: aa.molavich@ut.ac.ir

## I. Introduction

Terahertz (THz) radiation is a special part of the electromagnetic spectrum and has gained increasing attention for applications such as non-destructive imaging, molecular fingerprint spectroscopy, ultrafast wireless communication, biomedical diagnostics and security screening [1–4]. But the realisation of compact, effcient, broadband and spectrally tunable THz sources remains one of the key challenges in modern photonics and optoelectronics [5–7]. Two-dimensional (2D) materials belong to the emerging material platforms that attract special attention owing to their strong light–matter interaction, ultrafast carrier dynamics, reduced dielectric screening, and highly tunable electronic properties, which provide unprecedented opportunities for engineering nonlinear optical processes and THz emission [8, 9]. MXenes have emerged as an outstanding platform owing to their  conductivity, high carrier density, tunable electronic structure and strong plasmonic response in the THz and infrared spectral regions. Different from graphene, the electronic properties of these materials can be systematically engineered by the choice of transition-metal compositions and surface terminations (–O, –OH and –F), providing the opportunity to precisely control the carrier concentration, band structure, scattering dynamics, optical conductivity and plasmonic behaviour [10–12]. MXenes have been widely investigated in the THz absorbers, electromagnetic interference (EMI) shielding, metamaterials and biosensing fields [13–16]. Experimental and theoretical results demonstrate effcient THz absorption, plasmon enhanced electromagnetic response, high performance shielding, ultrafast carrier dynamics controlled by surface chemistry and layered crystal structure. Ultrafast THz spectroscopy has further revealed the decisive influence of surface terminations on transient conductivity, carrier relaxation, and plasmonic resonances, while engineered MXene metamaterials have enabled highly sensitive THz biosensing through strong field localization. These advances establish MXenes as one of the most promising platforms for passive THz technologies. Nevertheless, despite their exceptional electronic and optical properties, the active exploitation of MXenes as coherent THz emitters remains largely unexplored.

Meanwhile, considerable progress has been achieved in coherent THz generation from plasmas, graphene, and other 2D materials [17–20]. Theoretical studies based on tight-binding Hamiltonians, semiconductor Bloch equations, and density-matrix formalisms have established that intense femtosecond excitation can drive highly nonlinear interband and intraband carrier dynamics, giving rise to broadband THz radiation. These predictions have been supported experimentally by coherent THz emission, nonlinear photon-drag effects, and high-harmonic generation in graphene under ultrafast optical excitation. Particular attention has been devoted to two-color femtosecond fields for waveform control, where a controlled phase difference between the fundamental and second-harmonic components breaks inversion symmetry and generates a net photocurrent, leading to enhanced low-frequency THz emission [21]. More advanced quantum-kinetic studies have further demonstrated that the emission dynamics are governed not only by transient carrier populations but also by quantum coherence and the temporal evolution of the electronic density matrix, extending the physical description beyond simplified semiclassical and tunneling-based pictures [22, 23]. Recent investigations have expanded these concepts to multicolor optical fields, revealing additional routes for controlling photocurrents, high-harmonic generation, and the polarization and efficiency of the emitted radiation [24, 25]. All-optical quantum-control strategies have further demonstrated that tailored femtosecond waveforms can manipulate ultrafast carrier dynamics and thereby modify THz emission characteristics in both two-dimensional and bulk materials [26].

Despite these advances, a quantum-mechanical description capable of consistently explaining coherent THz generation in MXenes has yet to be established. Existing theoretical descriptions of THz emission in MXenes have predominantly relied on classical transport, Drude-type conductivity, or phenomenological hydrodynamic models, in which carrier relaxation is introduced phenomenologically and microscopic quantum coherence is not explicitly retained. As a result, the influence of surface terminations, electron–phonon interactions, impurity scattering, and nonequilibrium carrier redistribution remains diffcult to incorporate consistently within a unified microscopic description of the electromagnetic response. Moreover, quantum-kinetic approaches originally developed for graphene have only rarely been extended to MXenes, whose transition-metal d-orbitals and chemically tunable surface terminations generate substantially richer electronic structures and scattering pathways. Likewise, waveform engineering beyond conventional two-color excitation has received little attention for MXene-based THz emitters. A comprehensive theoretical description that consistently connects MXene-dependent effective electronic structure, quantum coherence, nonlinear carrier transport, environmentally induced relaxation, and coherent THz radiation has therefore remained unavailable. To address these challenges, a multiscale density-matrix formalism for coherent THz generation in MXenes driven by engineered three-color femtosecond waveforms is developed. MXene-dependent effective electronic structures are described using parameterized multiband tight-binding Hamiltonians, while ultrafast light–matter interaction is formulated within a density-matrix approach combined with a non-secular Redfield master equation that explicitly retains interband coherence and carrier populations, while environmental effects associated with electron–phonon interactions, impurity scattering, and surface termination are represented through effective relaxation and dephasing channels. Rather than relying exclusively on computationally intensive first-principles simulations, environmental effects are incorporated through physically motivated relaxation channels and bath-correlation functions, preserving the essential microscopic physics while enabling computationally effcient simulations over experimentally relevant parameter spaces. In contrast to previous theoretical approaches, the present methodology resolves the microscopic interplay between multiband quantum coherence, environmentally induced relaxation, nonlinear carrier transport, and coherent THz emission within a unified computational description. The resulting carrier dynamics are directly connected to the emitted THz field through the transient photocurrent and its time derivative, thereby establishing an explicit relationship between engineered optical waveforms, coherent quantum dynamics, microscopic carrier transport, and macroscopic THz radiation. Three-color waveform parameters are systematically explored to enable quantitative comparisons among different MXene compositions under identical excitation conditions. Beyond providing quantitative predictions of coherent THz emission, the present theory establishes a portable microscopic design platform for connecting surface chemistry, quantum coherence, open-system relaxation, nonlinear carrier transport, and electromagnetic radiation across MXenes and related low-dimensional quantum materials.

# II. Theoretical Framework

## 2.1 Effective Low-Energy Hamiltonian and Engineered Optical Waveforms

Coherent THz generation in MXenes is governed primarily by electronic states located near the Fermi level, where near-infrared femtosecond excitation induces ultrafast nonequilibrium carrier dynamics. Because deeper valence states contribute only weakly to the photoinduced response, the essential physics can be captured within an effective low-energy electronic model. Accordingly, the MXene monolayer is treated as an open quantum system, in which coherent light–matter interaction is described by an effective electronic Hamiltonian, while environment-induced dissipation arising from electron–phonon interactions, impurities, structural disorder, and surface functionalization is incorporated separately through the Redfield quantum master equation introduced in Sec. II.C. The coherent electronic Hamiltonian is decomposed as $\widehat{H}(t) = \widehat{H}_0 + \widehat{H}_{\text{int}}(t)$, where $\widehat{H}_0$ describes the field-free electronic structure and $\widehat{H}_{\text{int}}(t)$ represents the interaction with the time-dependent optical field:

$$\widehat{H}(t) = \widehat{H}_0 + \widehat{H}_{\text{int}}(t). \tag{1}$$

The electronic structure of the reference MXene is represented by an effective six-orbital tight-binding Hamiltonian in the basis $\{ |\varphi_1\rangle, \ldots, |\varphi_6\rangle \}$. The field-free Bloch Hamiltonian is written as, $H_0(k) = H_{\text{onsite}} + H_{nn}(k)$, where $[H_{\text{onsite}}]_{ij} = \varepsilon_i\delta_{ij}$ and $[H_{nn}(k)]_{ij} = \sum_{n=1}^{3} t_{ij}^{(n)} \exp(i\, k \cdot \delta_n)$. Here, $\delta_n$ denotes the nearest-neighbour bond vectors and $t_{ij}^{(n)}$ the corresponding hopping parameters. This effective model describes the relevant low-energy electronic states without attempting to reproduce the complete first-principles electronic structure. Rather than employing a fully atomistic first-principles Hamiltonian, an effective tight-binding representation is adopted to describe the low-energy electronic states participating in the ultrafast optical excitation. Within a localized Wannier basis $\{|i\rangle\}$, the field-free Hamiltonian is written as:

$$\hat{H}_0 = \sum_i \varepsilon_i |i\rangle\langle i| + \sum_{\langle i,j\rangle} t_{ij} |i\rangle\langle j|, \tag{2}$$

where $\varepsilon_i$ and $t_{ij}$ denote the effective onsite energies and nearest-neighbour hopping integrals, respectively. These parameters represent a phenomenological low-energy description of MXenes and are not intended to reproduce the complete first-principles electronic structure. For periodic crystals, the Hamiltonian is conveniently transformed into reciprocal space as $\hat{H}_0 = \sum_{\mathbf{k}} \hat{H}(\mathbf{k})$, where $\mathbf{k}$ is the crystal momentum in the first Brillouin zone, providing the natural framework for describing laser-driven carrier dynamics and evaluating the microscopic current. Light–matter interaction is formulated in the velocity gauge using the Peierls substitution, whereby the hopping amplitudes acquire the phase factor:

$$t_{ij} \to t_{ij} \exp\left[-\frac{ie}{\hbar}\int_{\mathbf{r}_i}^{\mathbf{r}_j} \mathbf{A}(t) \cdot d\mathbf{l}\right], \tag{3}$$

Here, $e > 0$ denotes the elementary charge, and the electron charge is $-e$, with $\mathbf{A}(t)$ denoting the vector potential of the incident optical waveform. This formulation preserves lattice translational symmetry and provides a consistent description of coherent carrier acceleration under strong optical excitation. MXene-dependent effective characteristics are incorporated primarily through the effective Hamiltonian parameters and environmental spectral densities, providing a portable theoretical framework that can be adapted to different MXene compositions with appropriate calibration of the corresponding electronic and relaxation parameters. Efficient coherent THz generation requires optical excitation with broken temporal symmetry. Under a temporally symmetric driving field, carrier acceleration during one half-cycle is compensated by the subsequent half-cycle, resulting in negligible residual photocurrent. Recent advances in ultrafast pulse shaping have made it possible to overcome this limitation by coherently synthesizing multiple frequency components with independently controlled amplitudes, temporal delays, and polarization states. Such waveform engineering enables direct manipulation of carrier acceleration, quantum interference, and nonequilibrium transport, thereby providing an effective route for controlling coherent THz emission. To retain full generality, based on Figure 1, the incident optical field is represented as:

$$\mathbf{E}(t) = \sum_{j=1}^{N} E_j f_j(t) \cos(\omega_j t + \delta_j) \hat{\mathbf{e}}_j, \tag{4}$$

where $E_j$, $\omega_j$, $\delta_j$, $\tau_j$, and $\hat{\mathbf{e}}_j$ denote the amplitude, carrier frequency, temporal delays, and polarization vector of the $j$th spectral component, respectively. Each component is assumed to possess a Gaussian envelope, $f_j(t) = \exp[-4\ln 2\,(t/\tau_{p,j})^2]$, where $\tau_{p,j}$ is the full width at half maximum (FWHM) pulse duration. This generalized formulation encompasses single-color, multicolor, and arbitrarily shaped excitation fields, while the numerical examples presented below focus on three mutually coherent femtosecond frequency components. The optical interaction is formulated in the velocity gauge, with the vector potential given by $\mathbf{A}(t) = -\int_{-\infty}^{t} \mathbf{E}(t')\, dt'$, which enters the Hamiltonian through the Peierls substitution introduced in Sec. II.A. Consequently, the temporal symmetry, spectral composition, and polarization states of the engineered waveform constitute the primary external control parameters governing coherent carrier acceleration, the competition between quantum coherence and environmental relaxation, and ultimately the efficiency and spectral characteristics of the emitted THz radiation.

### 2.2 Open Quantum Dynamics and Redfield Master Equation

Ultrafast carrier dynamics in MXenes are governed not only by coherent optical excitation but also by interactions with the surrounding environment. In the present implementation, these environmental effects are represented by effective phenomenological relaxation and dephasing channels rather than by explicitly separated microscopic scattering mechanisms. The corresponding reference bath parameters determine the effective relaxation and coherence-dephasing times used in the calculations. Accordingly, the laser-driven electronic subsystem is treated as the system of interest, while all remaining microscopic degrees of freedom are incorporated into an effective environment. The total Hamiltonian of the coupled system is written as:

$$\hat{H}_{\mathrm{tot}}(t) = \hat{H}(t) + \hat{H}_{\mathrm{B}} + \hat{H}_{\mathrm{SB}}, \tag{5}$$

where $\hat{H}(t)$ is the laser-driven electronic Hamiltonian introduced in Sec. II.A, $\hat{H}_{\mathrm{B}}$ describes the environmental bath, and $\hat{H}_{\mathrm{SB}}$ represents the system–bath interaction.

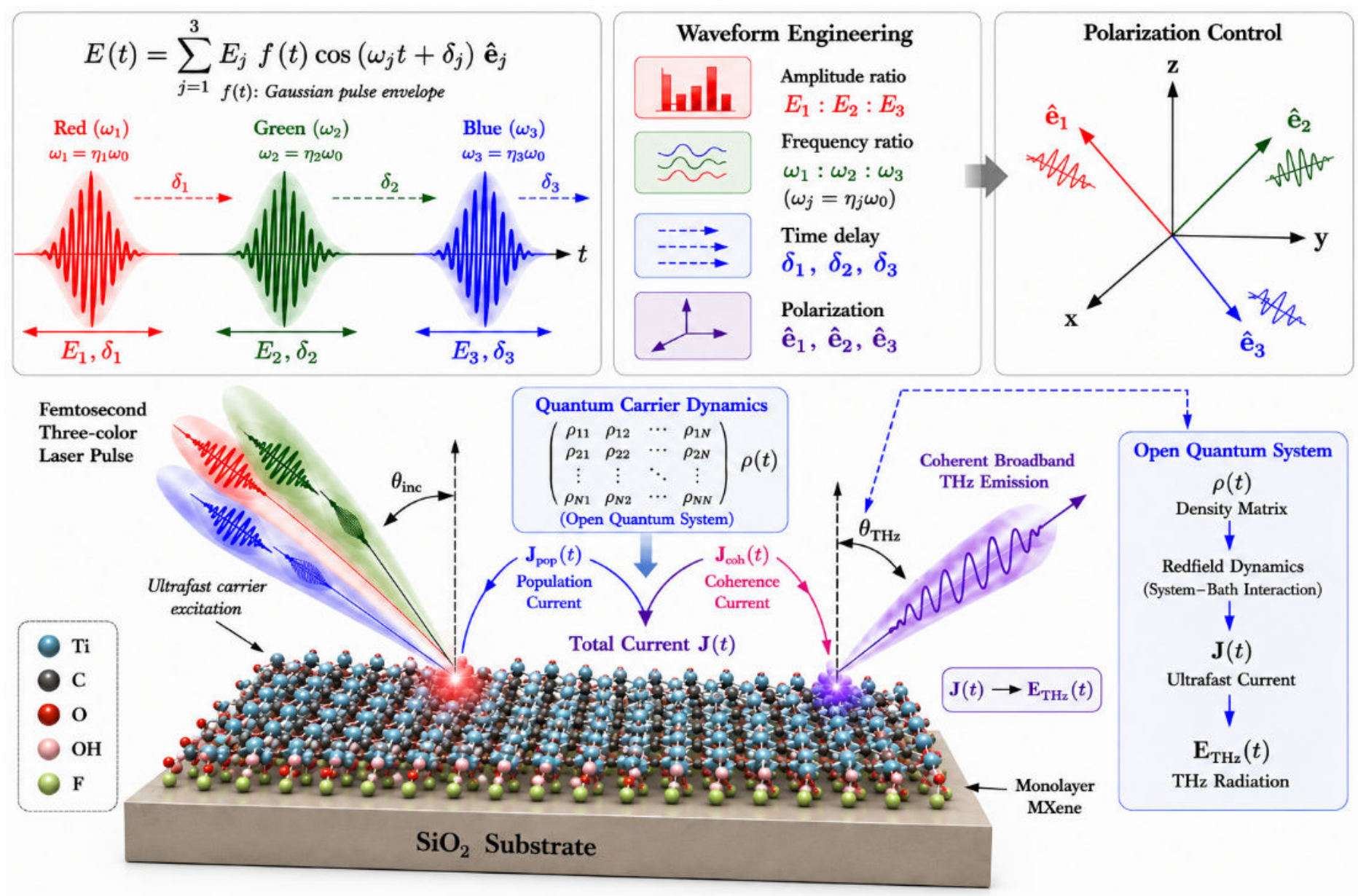

**Figure 1:** Conceptual illustration of the proposed multiscale framework for coherent THz generation from monolayer MXenes driven by an engineered three-color femtosecond waveform.

The environment is modeled as a collection of independent harmonic oscillators, $\hat{H}_{\mathrm{B}} = \sum_{\lambda} \hbar\omega_{\lambda}(b_{\lambda}^{\dagger} b_{\lambda} + 1/2)$, while the system–environment coupling is assumed to be bilinear, $\hat{H}_{\mathrm{SB}} = \sum_{\alpha} \hat{S}_{\alpha} \otimes \hat{B}_{\alpha}$, where $\hat{S}_{\alpha}$ and $\hat{B}_{\alpha}$ operate on the electronic and environmental Hilbert spaces, respectively. Different environmental relaxation mechanisms are incorporated phenomenologically through the associated bath correlation functions and spectral densities, allowing the framework to account for processes such as electron–phonon coupling, impurity-related scattering, and surface-functionalization effects. The dynamics of the complete system obey the Liouville–von Neumann equation:

$$\frac{d}{dt}\hat{\rho}_{\mathrm{tot}}(t) = -\frac{i}{\hbar}\left[\hat{H}_{\mathrm{tot}}(t), \hat{\rho}_{\mathrm{tot}}(t)\right], \tag{6}$$

where $\hat{\rho}_{\mathrm{tot}}(t)$ denotes the density operator of the coupled system. Since the experimentally accessible observables depend only on the electronic subsystem, the environmental degrees of freedom are eliminated by tracing over the bath, $\hat{\rho}(t) = \mathrm{Tr}_{\mathrm{B}}[\hat{\rho}_{\mathrm{tot}}(t)]$, yielding the reduced electronic density operator. The resulting reduced density matrix provides direct access to carrier populations, quantum coherences, microscopic photocurrents, and, ultimately, the emitted THz radiation. The remaining task is therefore to derive an effective equation of motion that consistently combines coherent light–matter interaction with irreversible environmental relaxation, as accomplished by the Redfield quantum master equation presented below. A direct solution of the Liouville–von Neumann equation is generally intractable because of the enormous number of environmental degrees of freedom. The dynamics are therefore transformed into the interaction picture with respect to the uncoupled Hamiltonian $\hat{H}_0(t) = \hat{H}(t) + \hat{H}_\mathrm{B}$, where the interaction-picture operators are generated by the time-evolution operator $U_0(t) = \mathcal{T}\exp[-(i/\hbar)\int_0^t \hat{H}_0(t')dt']$. The equation of motion then assumes the compact form:

$$\frac{d}{dt}\tilde{\rho}_\mathrm{tot}(t) = -\frac{i}{\hbar}\left[\tilde{H}_\mathrm{SB}(t), \tilde{\rho}_\mathrm{tot}(t)\right], \tag{7}$$

which provides the starting point for the Redfield formalism. Formal integration of Eq. (7) followed by substitution back into the equation of motion yields the exact second-order integro-differential equation:

$$\frac{d}{dt}\tilde{\rho}_\mathrm{tot}(t) = -\frac{1}{\hbar^2}\int_{t_0}^{t} dt' \left[\tilde{H}_\mathrm{SB}(t), \left[\tilde{H}_\mathrm{SB}(t'), \tilde{\rho}_\mathrm{tot}(t')\right]\right], \tag{8}$$

where an initially factorized state, $\hat{\rho}_\mathrm{tot}(0) = \hat{\rho}(0) \otimes \hat{\rho}_\mathrm{B}$, has been assumed. Under the Born approximation, the weak system–bath coupling allows the total density operator to remain approximately factorized throughout the evolution, $\tilde{\rho}_\mathrm{tot}(t) \approx \tilde{\rho}(t) \otimes \hat{\rho}_\mathrm{B}$. Tracing over the environmental degrees of freedom gives:

$$\frac{d}{dt}\tilde{\rho}(t) = -\frac{1}{\hbar^2}\int_{t_0}^{t} dt' \, \mathrm{Tr}_\mathrm{B}\left\{\left[\tilde{H}_\mathrm{SB}(t), \left[\tilde{H}_\mathrm{SB}(t'), \tilde{\rho}(t') \otimes \hat{\rho}_\mathrm{B}\right]\right]\right\}, \tag{9}$$

which remains non-Markovian because the evolution depends on the entire history of the reduced density operator. For condensed-matter environments, however, the bath correlation time is typically much shorter than the characteristic relaxation time of the electronic subsystem. Under this condition, the Markov approximation replaces $\tilde{\rho}(t')$ by $\tilde{\rho}(t)$ and extends the memory integral to infinity, yielding a time-local evolution equation. The influence of the environment is completely characterized by the bath correlation functions:

$$C_{\alpha\beta}(\tau) = < \tilde{B}_\alpha(\tau)\tilde{B}_\beta(0) >_\mathrm{B}, \tag{10}$$

whose Fourier transform,

$$J_{\alpha\beta}(\omega) = \int_{-\infty}^{\infty} C_{\alpha\beta}(\tau)e^{i\omega\tau}\, d\tau, \tag{11}$$

defines the environmental spectral density. In the present implementation, the environmental coupling is described by an effective phenomenological bath model. The bath correlation functions and spectral-density representation provide the formal basis of the Redfield dissipator, while the effective relaxation and dephasing parameters are specified through the reference bath rates used in the calculations. These rates represent experimentally motivated electronic relaxation and coherence-dephasing timescales rather than a unique microscopic scattering mechanism. These quantities enter the Redfield relaxation tensor derived in the following subsection. Under the Born–Markov approximation, the reduced dynamics become local in time and are governed by the Redfield quantum master equation. The coherent evolution is governed by the explicitly time-dependent Hamiltonian H(t), which contains the engineered optical waveform. In contrast, the dissipative Redfield tensor is constructed in the eigenbasis of the field-free Hamiltonian $H_0$ using the corresponding stationary Bohr frequencies. The effective relaxation and dephasing parameters are kept fixed during the driven evolution. Thus, the present calculations employ a field-free-basis non-secular Redfield dissipator coupled to the fully time-dependent coherent dynamics. This approximation retains the complete temporal dependence of the optical excitation while providing an effective stationary description of environmental relaxation and decoherence. Substituting the bath correlation functions into the reduced Born–Markov equation gives:

$$\frac{d}{dt}\hat{\rho}(t) = -\frac{i}{\hbar}[\hat{H}(t), \hat{\rho}(t)] + \mathcal{R}[\hat{\rho}(t)], \tag{12}$$

where the first term describes coherent laser-driven evolution and the Redfield superoperator $\mathcal{R}$ accounts for environmentally induced relaxation and decoherence. Expressed in the eigenbasis of the field-free Hamiltonian, $\hat{H}_0|m\rangle = E_m|m\rangle$, the reduced density matrix satisfies:

$$\frac{d\rho_{mn}(t)}{dt} = -\frac{i}{\hbar}\sum_r [H_{mr}(t)\rho_{rn}(t) - \rho_{mr}(t)H_{rn}(t)] + \sum_{pq} R_{mn,pq}\rho_{pq}(t), \tag{13}$$

where $H_{mn}(t) = E_m\delta_{mn} + H_{\text{int},mn}(t)$, and $R_{mn,pq}$ denotes the time-dependent Redfield relaxation tensor. The latter is given by:

$$R_{mn,pq} = \Gamma^{(+)}_{qn,mp} + \Gamma^{(-)}_{qn,mp} - \delta_{np}\sum_r \Gamma^{(+)}_{mr,rq} - \delta_{mq}\sum_r \Gamma^{(-)}_{nr,rp} \tag{14}$$

The transition rates are determined by the bath correlation functions:

$$\Gamma^{(+)}_{mn,pq} = \frac{1}{\hbar^2}\int_0^\infty C_{mn,pq}(\tau)e^{-i\omega_{pq}\tau}\, d\tau, \tag{15a}$$

$$\Gamma^{(-)}_{mn,pq} = \frac{1}{\hbar^2}\int_0^\infty C^*_{mn,pq}(\tau)e^{i\omega_{mn}\tau}\, d\tau. \tag{15b}$$

The Redfield tensor describes environmental relaxation and decoherence through the bath correlation functions, with the effective relaxation and dephasing rates specified by the reference bath parameters used in the present calculations. The resulting dissipative dynamics therefore provide an effective description of environmental relaxation and decoherence rather than a first-principles resolution of individual microscopic scattering channels. The present calculations employ the second-order non-secular Redfield description in the regime of weak system–bath coupling and rapidly decaying bath correlations, consistent with the Born–Markov approximation. The nonequilibrium carrier dynamics governed by the Redfield quantum master equation generate a transient microscopic photocurrent that acts as the fundamental source of the emitted THz radiation. Once the reduced density matrix has been obtained from Eq. (12), all electronic observables follow directly from the expectation values of the corresponding quantum operators. Among them, the photocurrent provides the essential bridge between microscopic quantum dynamics and macroscopic electromagnetic radiation. The microscopic current density is defined as $\mathbf{J}(t) = \text{Tr}[\hat{\rho}(t)\hat{\mathbf{J}}]$, where $\hat{\mathbf{J}}$ is the electronic current operator. For a periodic crystal, the current operator is related to the band velocity through $\hat{\mathbf{J}} = -(e/\hbar V)\nabla_{\mathbf{k}}\hat{H}(\mathbf{k}, t)$, where $V$ denotes the normalization volume (or crystal area for 2D systems). The total transient current is obtained by averaging the k-resolved current over the uniformly sampled first Brillouin zone. Accordingly, the microscopic current can be written as:

$$\mathbf{J}(t) = -\frac{e}{\hbar V}\text{Tr}[\hat{\rho}(t)\nabla_{\mathbf{k}}\hat{H}(\mathbf{k}, t)], \tag{16}$$

which establishes the direct connection between the nonequilibrium quantum state and the measurable electrical current. Expressed in the eigenbasis of the field-free Hamiltonian:

$$\mathbf{J}(t) = -\frac{e}{V}\sum_{mn} \rho_{nm}(t)\mathbf{v}_{mn}, \tag{17}$$

where $\mathbf{v}_{mn} = \langle m|\hat{\mathbf{v}}|n\rangle$ are the velocity matrix elements. Expressed in the eigenbasis of the field-free Hamiltonian, the transient photocurrent is determined by the velocity matrix elements weighted by the corresponding density-matrix elements. Equation (17) therefore reveals contributions from both diagonal density-matrix elements, representing carrier populations, and off-diagonal elements, representing quantum coherences. The population current reflects the nonequilibrium carrier distribution, whereas the coherent current originates from phase correlations established by the engineered optical waveform. Because these coherences are continuously modified by the Redfield relaxation tensor, the photocurrent naturally captures the competition between coherent optical control and environmentally induced decoherence. Since THz emission is generated exclusively by the photoinduced current, the equilibrium contribution is subtracted from the total current to obtain the transient photocurrent, $\Delta\mathbf{J}(t) = \mathbf{J}(t) - \mathbf{J}_{\text{eq}}$, where $\mathbf{J}_{\text{eq}} = \text{Tr}[\hat{\rho}_{\text{eq}}\ \hat{\mathbf{J}}]$ denotes the equilibrium current. The resulting transient photocurrent serves as the microscopic source term for the coupled quantum hydrodynamic and Maxwell equations developed in the following sections, thereby providing the direct physical link between quantum carrier dynamics and coherent THz radiation. The population–coherence decomposition is defined in the eigenbasis of the field-free Hamiltonian $H_0$ and is therefore basis dependent. Within this representation, the diagonal density-matrix elements define the population contribution, while the off-diagonal elements define the coherence contribution, such that the total transient current is exactly recovered as the sum of the two components.

### 2.3 Electromagnetic Radiation and THz Emission

The transient photocurrent obtained from the quantum master equation acts as the microscopic source of the emitted electromagnetic radiation, thereby establishing the connection between the nonequilibrium carrier dynamics and the experimentally observable THz signal. As described by Eqs. (16) and (17), this current is determined from the velocity matrix elements weighted by the nonequilibrium density-matrix elements, with contributions from both carrier populations and quantum coherences. Once the equilibrium contribution is subtracted, the resulting transient photocurrent, $\Delta\mathbf{J}(t)$, provides the source term for the electromagnetic radiation. The emitted electric field is obtained by relating this transient current source to the electromagnetic field through Maxwell's equations. Combining Maxwell's equations gives the inhomogeneous electromagnetic wave equation:

$$\nabla^2\mathbf{E} - \frac{1}{c^2}\frac{\partial^2\mathbf{E}}{\partial t^2} = \mu_0\frac{\partial\mathbf{J}}{\partial t}, \tag{18}$$

Here, $\mathbf{J}$ denotes the current density associated with the photoinduced electronic dynamics. For the ultrathin MXene system considered here, the spatially localized transient current is treated within the electric-dipole approximation. In this approximation, the radiating source is represented by an effective electric dipole, and the far-field radiation is proportional to the second time derivative of the dipole moment, or equivalently to the time derivative of the corresponding transient current. The emitted THz electric field is therefore written as:

$$\mathbf{E}_{\text{T}Hz}(t) \propto \frac{d}{dt}\Delta\mathbf{J}(t), \tag{19}$$

where $\Delta\mathbf{J}(t)$ denotes the transient photoinduced current obtained after subtracting the equilibrium current. Equation (19) provides the effective current-to-radiation relation adopted in the present framework. The proportionality factor contains the geometrical and propagation-dependent factors associated with the dipole radiation, which are not explicitly evaluated here because the analysis focuses on the normalized THz waveform and its relative spectral response. In the frequency domain, Eq. (19) becomes $\mathbf{E}_{\text{T}Hz}(\omega) = i\omega\mathbf{J}(\omega)$, where $\mathbf{J}(\omega)$ is the Fourier transform of the transient photoinduced current. Accordingly, the corresponding THz spectral intensity is obtained from

$$I_{\text{T}Hz}(\omega) \propto |\mathbf{E}_{\text{T}Hz}(\omega)|^2 \tag{20}$$

This relation provides the spectral representation of the calculated THz electric field. Because the transient photocurrent is obtained directly from the Redfield density-matrix dynamics, the resulting THz waveform and spectrum reflect the combined effects of coherent optical excitation, carrier acceleration, quantum coherence, and environment-induced relaxation. Variations in the amplitudes, temporal delays, and polarization states of the engineered optical waveform modify the density-matrix evolution and thereby control the transient photocurrent and the emitted THz radiation.

The physical basis of the present description is grounded in experimentally established ultrafast and THz properties of $Ti_3C_2T_x$ MXenes, with supporting context from related 2Dmaterials. THz time-domain spectroscopy has demonstrated strong free-carrier responses and ultrafast carrier scattering in $Ti_3C_2T_x$ films, providing experimentally informed ranges for carrier density, conductivity and momentum-relaxation times [27–29]. Femtosecond nonlinear-optical measurements have further established pronounced nonlinear optical responses in few-layer $Ti_3C_2T_x$ across the near-infrared [30], while experimental and first-principles studies have shown that surface terminations can substantially modify the electronic structure, carrier dynamics, and THz optical response of MXenes [30, 31]. Ultrafast optical and THz measurements have also revealed coupled electronic, plasmonic, and phononic dynamics on femtosecond-to-picosecond timescales [32, 33], providing a physical basis for incorporating environment-induced relaxation into microscopic descriptions of MXene carrier dynamics. These experimental and theoretical results do not constitute a direct validation of the present calculations, which address a state-resolved microscopic description of coherent THz generation under engineered multi-colour excitation. Rather, they provide experimentally grounded constraints and physical motivation for the material parameters, relaxation channels, nonlinear response, and THz frequency range explored here. Experimental studies of related 2Dquantum materials have further established coherent THz emission driven by ultrafast photocurrents, as demonstrated in graphene [34], layered transition-metal dichalcogenides [35, 36], and Weyl semimetals [37]. These observations support the physical connection between transient microscopic currents and emitted THz radiation that underlies the present description, while leaving open the experimental realization of analogous coherent emission from MXenes under engineered excitation.

The present approach extends the collective-carrier description of hydrodynamic–Maxwell models toward a stateresolved microscopic treatment by explicitly incorporating multiband electronic structure, interband quantum coherence, open-system relaxation, and electromagnetic radiation within a common computational scheme. Unlike semiclassical Boltzmann or rate-equation approaches, which retain only band populations, the density-matrix formalism preserves interband coherences and enables a direct decomposition of the transient current into population and coherence-driven components, $J(t) = J_{pop}(t) + J_{coh}(t)$. Experimental studies of 2Dquantum materials have directly linked nonlinear photocurrents to THz emission and demonstrated their sensitivity to electronic structure and optical driving conditions [38–41]. Semiconductor Bloch equations likewise provide a microscopic description of interband polarization and light–matter coupling, whereas relaxation is commonly introduced through phenomenological $T_1$ and $T_2$ times [42, 43]. Here, relaxation is incorporated through a non-secular Redfield tensor, allowing populations and coherences to evolve self-consistently within a common system–bath description. Benchmark comparisons of full Redfield dynamics with its secular approximation highlight the role of non-secular terms in the treatment of coupled electronic coherences [44–46]. Such a treatment is particularly relevant under broadband multi-colour excitation involving multiple nearly resonant interband transitions, where coherence transfer between coupled excitation pathways can contribute to the transient current and emitted THz response. Time-dependent density functional theory (RT-TDDFT) provides an atomistic first-principles description of ultrafast electronic dynamics and has been developed for strong-field and pump–probe simulations [47, 48]. Its computational cost, however, can constrain systematic exploration of waveform parameters, coherence lifetimes, environmental coupling, and material-design variables across multidimensional parameter spaces, while the controlled incorporation of explicit open-system relaxation introduces additional methodological complexity [47, 48]. Lindblad master equations provide a widely used route for open-quantum-system dynamics and can accommodate coupling to structured electromagnetic environments [49]. The present non-secular Redfield treatment instead retains the coherence-transfer terms associated with coupled transitions within the underlying system–bath dynamics [46]. The analyses presented above further show that environmental interactions do not merely attenuate the emitted THz signal. Instead, they redistribute the temporal evolution of electronic coherence, modify the relative balance between population- and coherence-driven currents, and reshape the emitted THz waveform in both the time and frequency domains. Taken together, these results establish a direct microscopic connection between MXene-dependent effective electronic structure, quantum coherence, open-system relaxation, nonlinear carrier transport, and coherent THz radiation, providing a physically transparent and computationally effcient foundation for waveform engineering and materials design of broadband coherent THz emitters.

## III. Results & Discussion

The proposed multiscale methodology was implemented through a coupled computational workflow linking the effective multiband electronic Hamiltonian, engineered optical excitation, open-quantum-system dynamics, microscopic photocurrent generation, and electromagnetic wave propagation. The calculation begins by constructing the effective Hamiltonian on a uniform sampling of the first Brillouin zone, from which the electronic eigenstates and velocity matrix elements are obtained. The interaction with the engineered optical waveform is incorporated through the Peierls substitution, yielding the time-dependent The reduced density matrix is initialized in thermal equilibrium at T = 300 K using Fermi–Dirac occupations, with the energy zero chosen at the chemical potential ($\mu = 0$ eV). The initial equilibrium state is therefore defined by the finite-temperature electronic occupation of the field-free Hamiltonian $H_0$ and is subsequently propagated by numerically solving the non-secular Redfield master equation using a fourth-order Runge–Kutta (RK4) scheme. Throughout the propagation, the Hermiticity and normalization of the density matrix are monitored to ensure numerical stability and physical consistency. To assess the physical consistency of the non-secular Redfield evolution, the eigenvalues of the reduced density matrix were monitored throughout the numerical propagation. The density matrix remained positive semidefinite within numerical precision, while trace preservation and Hermiticity were maintained throughout the simulations. At each time step, the transient photocurrent is evaluated from the instantaneous density matrix and subsequently coupled to the electromagnetic solver to determine the emitted THz field and its corresponding spectrum. The procedure is repeated for different excitation configurations by systematically varying the excitation intensity, relative amplitudes of the three spectral components, polarization configuration, inter-pulse temporal delays, and laser incidence angle. Combined variations of spectral amplitudes and excitation intensity are also considered to assess the response across the coupled waveform-parameter space, enabling direct comparison of the THz emission under controlled computational conditions. Numerical convergence is verified with respect to Brillouin-zone sampling, temporal discretization, and total propagation time. These parameters are refined until the calculated carrier dynamics, transient photocurrent, emitted THz waveform, and spectral characteristics remain invariant within the prescribed numerical tolerance. The reference calculations employ a six-orbital tight-binding Hamiltonian on a hexagonal $\mathrm{Ti}_3C_2T_x$ lattice with $a = 3.05$ Å. Onsite energies $\{-1.20, -0.85, -0.55, 0.45, 0.90, 1.35\}$ eV and hopping scale $t_0 = 0.55$ eV define the effective multiband electronic structure used to resolve the valence and conduction manifolds; these are model parameters rather than direct experimental fits. The carrier dynamics are integrated using a fourth-order Runge–Kutta scheme for the non-secular Redfield equation over $t \in [-100,100]$ fs with $\Delta t = 0.01$ fs. The reduced dynamics are initialized at a general initial time $t_0$. In the numerical implementation, $t_0 = -100$ fs is used, such that the system is initialized in thermal equilibrium before the arrival of the optical pulse. The reference bath rates $\gamma_{\mathrm{rel}}$ and $\gamma_{\mathrm{deph}}$ define the effective population-relaxation and coherence-dephasing times, $T_1 = 1/\gamma_{\mathrm{rel}}$ and $T_2 = 1/\gamma_{\mathrm{deph}}$, and are $\gamma_{\mathrm{deph}} = 1/(200\ \mathrm{fs})$ and $\gamma_{\mathrm{rel}} = 1/(600\ \mathrm{fs})$, respectively at $T = 300$ K. These timescales are motivated by experimentally established femtosecond-to-picosecond electronic, plasmonic, and electron–phonon dynamics in $\mathrm{Ti}_3C_2T_x$ [30, 7, 45], and are not identified with the much shorter momentum-scattering times of equilibrium Drude analyses. Optical excitation is a three-colour waveform at $\lambda_0 = 800nm$ $\{\omega_0, 2\omega_0, 3\omega_0\}$, amplitude ratios $E_2/E_1 = 0.50$ and $E_3/E_1 = 0.25$, fundamental peak field $E_1 = 3.5 \times 10^8$ V m $^{-1}$, and a common Gaussian envelope with an FWHM of 20 fs. The 800-nm drive and free-carrier/nonlinear optical response of $\mathrm{Ti}_3C_2T_x$ are consistent with established ultrafast optical and THz measurements [24, 30, 29, 23]. Unless stated otherwise, this reference set is used throughout; waveform, electronic-structure, and open-system variations are introduced explicitly in the corresponding analyses.

Engineering the excitation waveform provides an effective route for controlling the nonequilibrium carrier dynamics responsible for THz emission. Within the present quantum–hydrodynamic framework, the excitation intensity, spectral composition, polarization, relative temporal delay, and laser incidence angle govern the evolution of the reduced density matrix and reshape the transient photocurrent responsible for the emitted THz field. Increasing the excitation intensity enhances carrier acceleration and strengthens the transient photocurrent, leading to a larger THz response. The nonlinear variation of the peak field indicates that THz emission is governed by the coupled evolution of carrier populations and quantum coherences rather than by a simple linear scaling with the optical field. Redistributing the optical energy among the fundamental, second-, and third-harmonic components modifies the THz waveform by changing the coherent interference responsible for transient photocurrent formation. The spectral composition governs the temporal evolution of the microscopic photocurrent and the emitted THz radiation. Representative THz waveforms for the principal excitation schemes are compared in Supplementary Fig. S1, illustrating the distinct temporal signatures produced by single-, two-, and three-color excitation while emphasizing the robustness of the waveform-dependent emission dynamics. The polarization state changes both the amplitude and temporal profile of the THz field through its coupling to the electronic velocity matrix elements, highlighting the vectorial nature of coherent photocurrent generation. Femtosecond-scale temporal delays shift the dominant current burst, demonstrating that sub-cycle phase control provides an additional degree of freedom for tailoring the emission dynamics. The differential waveforms presented in Supplementary Fig. S2 further isolate the influence of envelope-delay engineering by removing the reference response, confirming that femtosecond-scale delays primarily redistribute the temporal structure of the transient photocurrent rather than introducing qualitatively different emission mechanisms. Varying the laser incidence angle with respect to the surface normal systematically modifies both the amplitude and temporal profile of the emitted THz waveform. Normal incidence ($\theta = 0°$) produces the strongest response, whereas increasing the incidence angle progressively reduces the emitted field. This angular dependence reflects the modification of the projected driving field and the associated carrier-acceleration dynamics under different excitation geometries. The results therefore demonstrate that the incidence angle provides an additional parameter for controlling coherent THz emission. The cumulative THz energy for the different waveform-engineering strategies is presented in Supplementary Fig. S3. The accumulated energy saturates after the optical interaction, while both the accumulation rate and the final emitted energy remain waveform dependent. Waveform engineering regulates THz emission by reorganizing the temporal evolution of nonequilibrium carrier dynamics responsible for transient photocurrent formation rather than by simply increasing the absorbed optical energy.

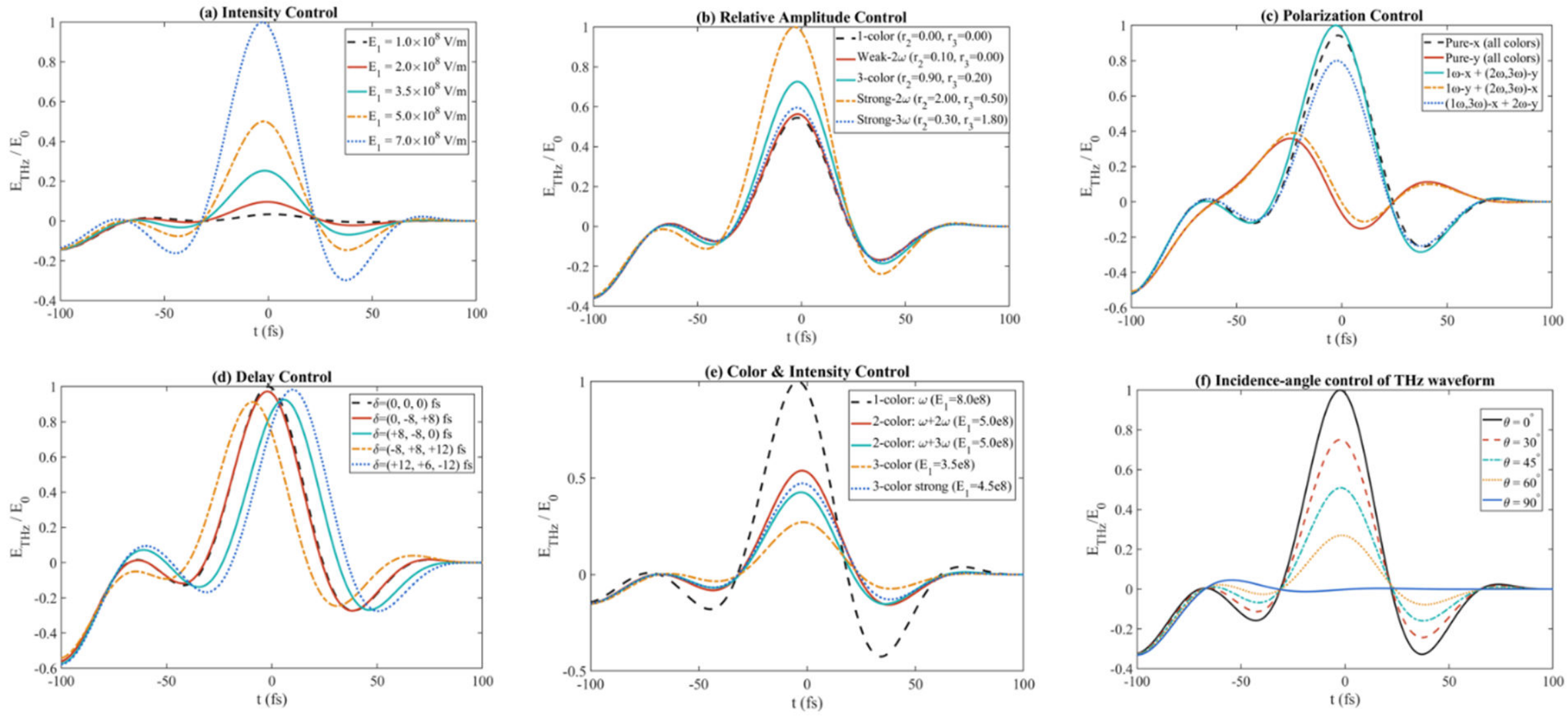

**Figure 2:** Systematic control of coherent THz emission through engineered three-color femtosecond waveforms. Normalized THz electric fields obtained under variations of (a) excitation intensity, (b) relative amplitudes of the three spectral components, (c) polarization configuration, (d) inter-pulse temporal delays, and (e) combined waveform engineering through systematic variation of spectral amplitudes and excitation intensity. Panel (f) presents the influence of the laser incidence angle, measured with respect to the surface normal, demonstrating the dependence of the emitted THz waveform on the excitation geometry.

Ultrafast THz generation originates from the coupled evolution of carrier populations and quantum coherences induced by optical excitation. Within the Redfield density-matrix framework, the optical field simultaneously redistributes electronic populations and establishes interband coherences, both contributing to the transient photocurrent that drives the emitted THz field. The results show that these processes evolve on comparable femtosecond timescales while playing distinct roles in the emission mechanism. The population dynamics reveal that optical excitation is dominated by charge transfer from the lowest valence state (VB1) to the first conduction state (CB1), whereas the remaining bands undergo only minor population changes. This behavior identifies the VB1–CB1 transition as the primary optically active channel governing nonequilibrium carrier generation under the present excitation conditions. The dominant population-transfer pathway does not necessarily coincide with the dominant radiating-current pathway. Population analysis identifies the transitions that contribute most strongly to nonequilibrium carrier generation, whereas the transition-resolved current is additionally weighted by the corresponding velocity matrix elements and density-matrix coherences. Consequently, a transition with a smaller population transfer can produce a larger photocurrent and radiated THz field when its current matrix elements and coherent response are stronger. The coherence dynamics are dominated by the VB1–CB1 interband coherence, whereas the remaining off-diagonal density-matrix elements exhibit substantially smaller amplitudes. The residual coherence after the pulse reflects finite dephasing within the Redfield description, indicating that phase coherence persists beyond the optical interaction over the characteristic relaxation times of the system. The density-matrix snapshots directly visualize this evolution. Before excitation, the electronic population is localized in the initial valence state with negligible coherence. Near the pulse maximum, population redistribution is accompanied by the development of interband coherence, demonstrating the simultaneous formation of nonequilibrium carriers and electronic polarization. After the pulse, the populations approach a quasi-stationary distribution while a finite coherent component remains. Together, these observations demonstrate that transient carrier redistribution and interband coherence play complementary roles in establishing the microscopic photocurrent responsible for THz emission, indicating that the emitted THz field arises from their coupled evolution rather than from population transfer alone. Additional microscopic evidence is provided in Supplementary Figs. S3 and S4, where decomposition of the transient photocurrent and evolution of the full density matrix reveal how carrier populations and quantum coherences jointly establish the ultrafast current responsible for THz emission.

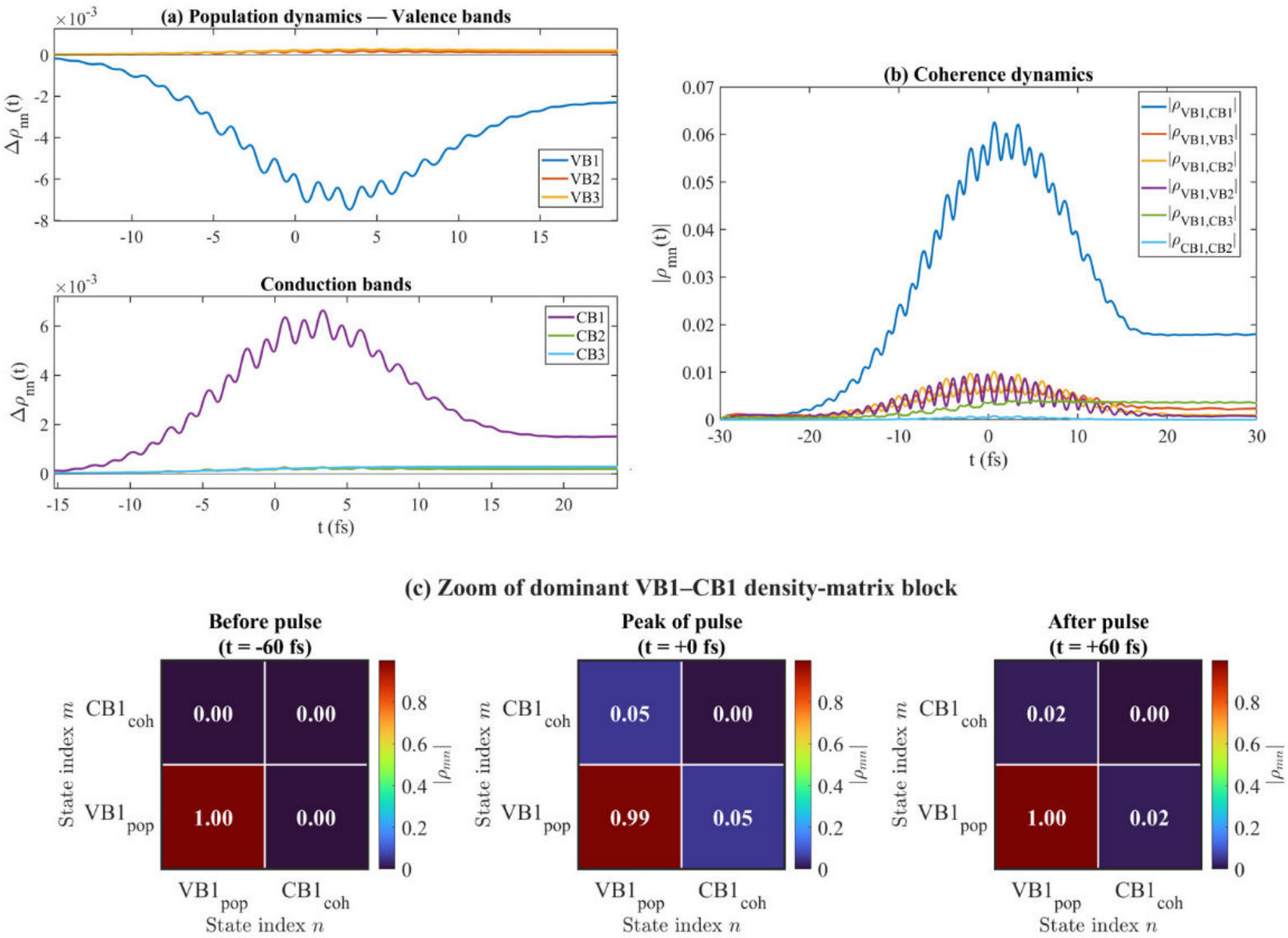

**Figure 3:** Ultrafast evolution of carrier populations and quantum coherences under three-color excitation. Time-dependent density-matrix dynamics obtained from the non-secular Redfield master equation. (a) Population dynamics in the valence and conduction bands, highlighting the dominant carrier transfer between VB1 and CB1. (b) Evolution of representative interband coherence amplitudes, showing the buildup of optical coherence during excitation followed by environment-induced relaxation. (c) Density-matrix snapshots of the VB1–CB1 subspace before, during, and after pulse excitation, illustrating the transient coexistence of population redistribution and interband coherence underlying ultrafast photocurrent generation.

The nonlinear THz response under single- and multi-harmonic excitation is characterized through the integrated THz energy, the local power-law exponent, the peak conduction-band population, and the corresponding nonlinear phase diagram is depicted in Fig. 4. The local exponent is used here as a scaling diagnostic of the integrated THz energy with respect to the driving field, rather than as a direct measure of the optical nonlinear order. Its field dependence provides a quantitative indicator of the evolving THz emission response and is used to identify the crossover from predominantly perturbative to increasingly nonlinear carrier dynamics. The nonlinear response of THz emission is governed by the spectral structure of the driving optical waveform. Rather than acting as a simple intensity enhancement mechanism, multi-harmonic excitation modifies how the electronic system evolves from the perturbative regime toward strong-field carrier dynamics. The field dependence of THz emission reflects changes in the underlying microscopic transport processes rather than merely variations in the excitation strength. At low optical fields, all excitation schemes exhibit comparable scaling exponents, indicating that carrier excitation remains predominantly perturbative. As the driving field increases, the local nonlinear exponent progressively rises, revealing the gradual onset of strong-field carrier dynamics. This continuous evolution demonstrates that the nonlinear transport regime develops smoothly with increasing excitation rather than through an abrupt transition. The simultaneous growth of the conduction-band population correlates with the enhanced THz response, indicating increasingly efficient nonequilibrium carrier excitation under stronger optical driving. This correlation indicates that the enhanced THz response accompanies the evolving nonequilibrium carrier dynamics established during optical excitation, consistent with the transient photocurrent predicted by the density-matrix framework. The nonlinear phase diagram shows that harmonic waveform engineering reshapes the boundary between perturbative and strong-field transport regimes. By reshaping nonlinear carrier dynamics, spectral waveform synthesis influences both the efficiency of THz generation and the microscopic pathway through which the transient photocurrent is established. These results demonstrate that multi-harmonic excitation provides a means of controlling the nonlinear transport regime itself, rather than simply increasing the emitted THz intensity.

The microscopic origin of THz emission is resolved by decomposing the transient photocurrent into individual interband transition channels in Fig. 5. Within the density-matrix framework, the contribution of an interband pair m → n is defined as $J_{mn}(t) = 2\ \mathrm{Re}[\rho_{nm}(t) j_{mn}(t)]$, where $\rho_{nm}(t)$ and $j_{mn}(t)$ are the corresponding density-matrix and current-operator matrix elements in the energy basis. The relative weight of each channel is quantified by the normalized current-power metric $P_{mn} = \int |J_{mn}(t)|^2 dt\ /\ \Sigma_{mn} \int |J_{mn}(t)|^2 dt$, providing a consistent measure for identifying the dominant transition pathways. A clear hierarchy emerges among the optically active transitions. The VB3→CB1 channel rapidly becomes the primary contributor to the emitted field, VB3→CB2 provides a smaller contribution, and the remaining transitions remain negligible throughout the excitation. THz emission is therefore dominated by a limited number of electronically active pathways rather than by the entire manifold of interband excitations. The transition-resolved photocurrents exhibit the same hierarchy. The VB3→CB1 channel carries the largest transition-resolved current, whereas the secondary channels generate substantially weaker currents with distinct temporal profiles. The dominant population-transfer pathway does not necessarily coincide with the dominant radiating-current pathway: population analysis characterizes electronic occupation redistribution, whereas the transition-resolved current is additionally determined by the corresponding density-matrix coherence and current matrix element. Thus, a transition with a smaller population transfer can nevertheless provide a stronger contribution to the emitted THz field. The transition-resolved spectra further reveal distinct spectral fingerprints, characterized by different peak frequencies and bandwidths. The calculated THz spectrum therefore reflects the superposition of spectrally distinct microscopic emission channels, shaped by the electronic dispersion and optical transition matrix elements associated with each pathway. The transition-pathway diagram summarizes the microscopic hierarchy identified by the density-matrix analysis. These dominant interband pathways determine the amplitude and spectral characteristics of the emitted THz radiation through their contributions to the transient photocurrent, providing a microscopic basis for tailoring coherent THz emission through controlled excitation of specific electronic channels.

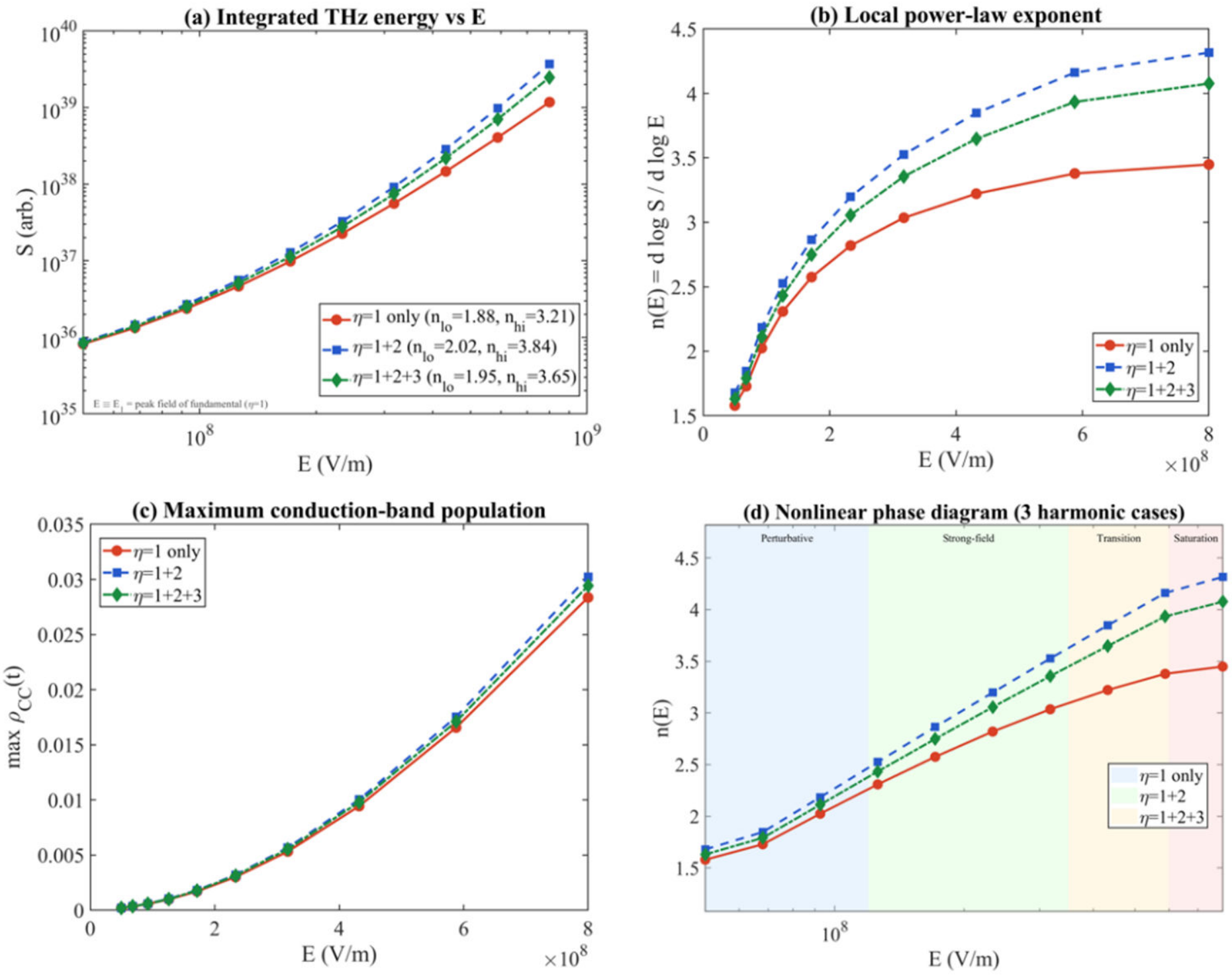

**Figure 4:** Nonlinear scaling of THz emission under multi-harmonic waveform engineering. (a) Integrated THz emission energy as a function of the fundamental peak field for three excitation configurations: the fundamental alone, the fundamental plus second harmonic, and the full three-harmonic waveform. The corresponding low- and high-field scaling exponents are indicated for each case. (b) Local nonlinear exponent, $n(E) = d\ln S/d\ln E$, revealing the continuous transition from the perturbative to the nonperturbative regime. (c) Peak conduction-band population as a function of driving field, demonstrating the progressive increase of photoexcited carriers with increasing excitation strength. (d) Nonlinear phase diagram summarizing the field-dependent operating regimes and illustrating how multi-harmonic waveform engineering shifts the onset and evolution of strong-field THz generation.

The influence of open-system dynamics on ultrafast THz generation is examined through the emitted THz waveforms, the corresponding frequency spectra, the evolution of quantum-state purity, and the relaxation–dephasing phase diagram (Fig. 6). Within the Redfield density-matrix framework, energy relaxation and pure dephasing modify the evolution of the reduced density matrix, thereby reshaping the transient photocurrent responsible for THz radiation. The calculated time-domain THz waveforms demonstrate that system–environment coupling systematically modifies both the amplitude and temporal profile of the emitted field. Concurrently, the spectra exhibit a redistribution of spectral weight toward lower frequencies together with a reduction in spectral bandwidth. These changes indicate that relaxation suppresses the faster temporal components of the microscopic photocurrent while preserving the ultrafast emission process. The purity of the reduced density matrix provides a measure of the mixedness of the electronic state, while its reduction under system–environment coupling is accompanied by a suppression of the coherent contribution to the photocurrent. Although the electronic state remains highly coherent throughout the interaction, stronger system-environment coupling produces a progressive reduction in purity, accompanied by an enhanced purity-loss rate concentrated near the peak of the driving pulse. These observations indicate that decoherence develops concurrently with nonequilibrium carrier excitation rather than emerging only after the optical interaction has ended. The relaxation-dephasing phase diagram shows that the normalized THz yield is more sensitive to the population relaxation time ($T_1$) than to the pure-dephasing time ($T_2$) within the investigated parameter range. Although the waveforms and spectra are normalized for shape comparison, the THz emission yield is evaluated independently from the corresponding unnormalized spectral intensity integrated over the specified THz frequency window. The THz generation efficiency is therefore primarily controlled by the lifetime of nonequilibrium carrier populations, while coherence decay has a comparatively weaker influence. These results identify relaxation and dephasing as complementary parameters governing the transient photocurrent and, consequently, the ultrafast THz emission in open quantum systems.

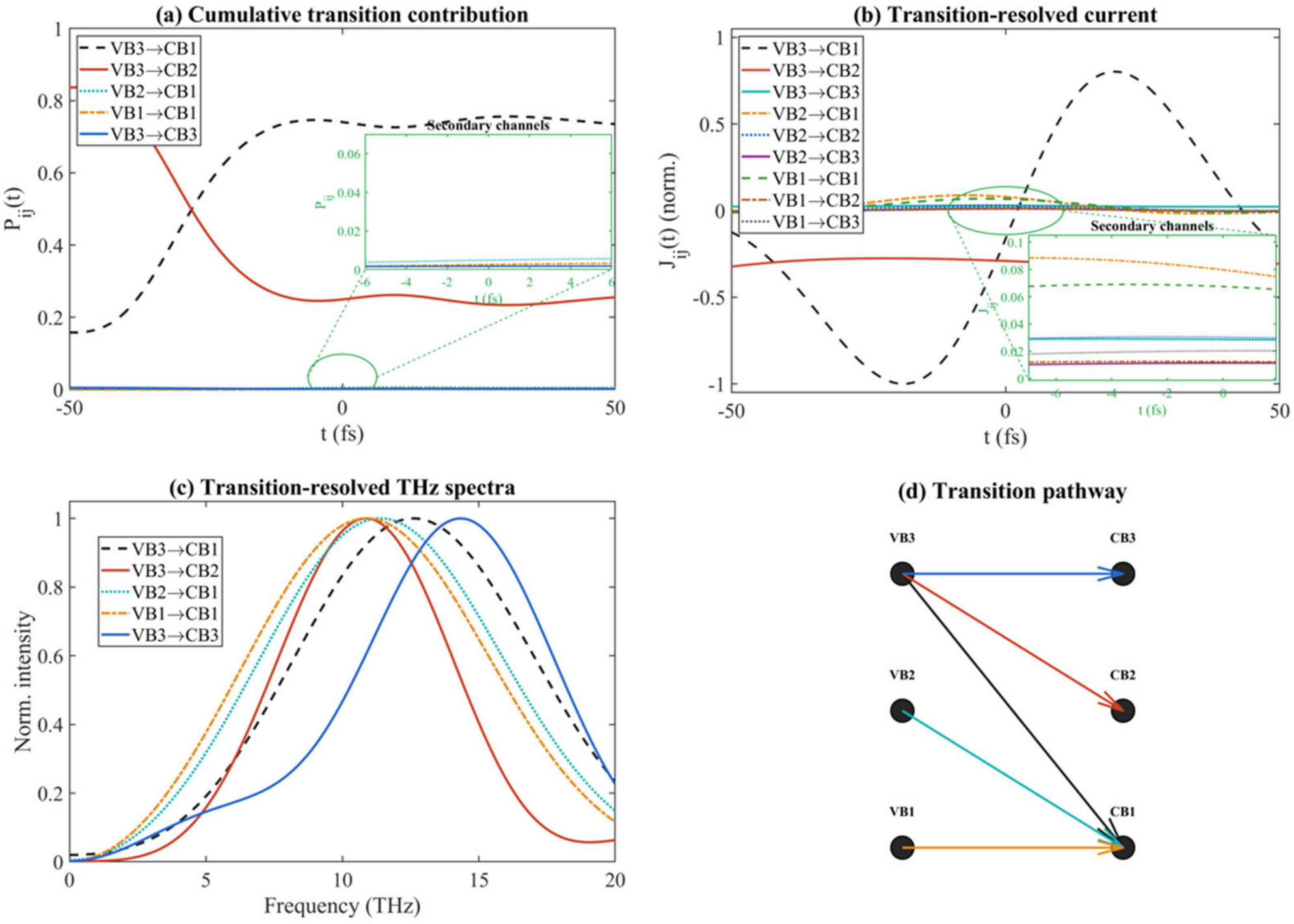

**Figure 5:** Transition-resolved microscopic pathways underlying coherent THz generation. (a) Time-dependent cumulative contributions of the dominant interband transitions to the emitted THz signal, illustrating the progressive emergence of the principal excitation channels. (b) Transition-resolved photocurrent contributions associated with the individual interband transitions, revealing their microscopic contributions to the transient current. The inset highlights the weaker secondary channels. (c) THz spectra associated with the dominant transition pathways, demonstrating their distinct spectral characteristics and emission bandwidths. (d) Schematic summary of the dominant interband excitation pathways identified from the density-matrix analysis.

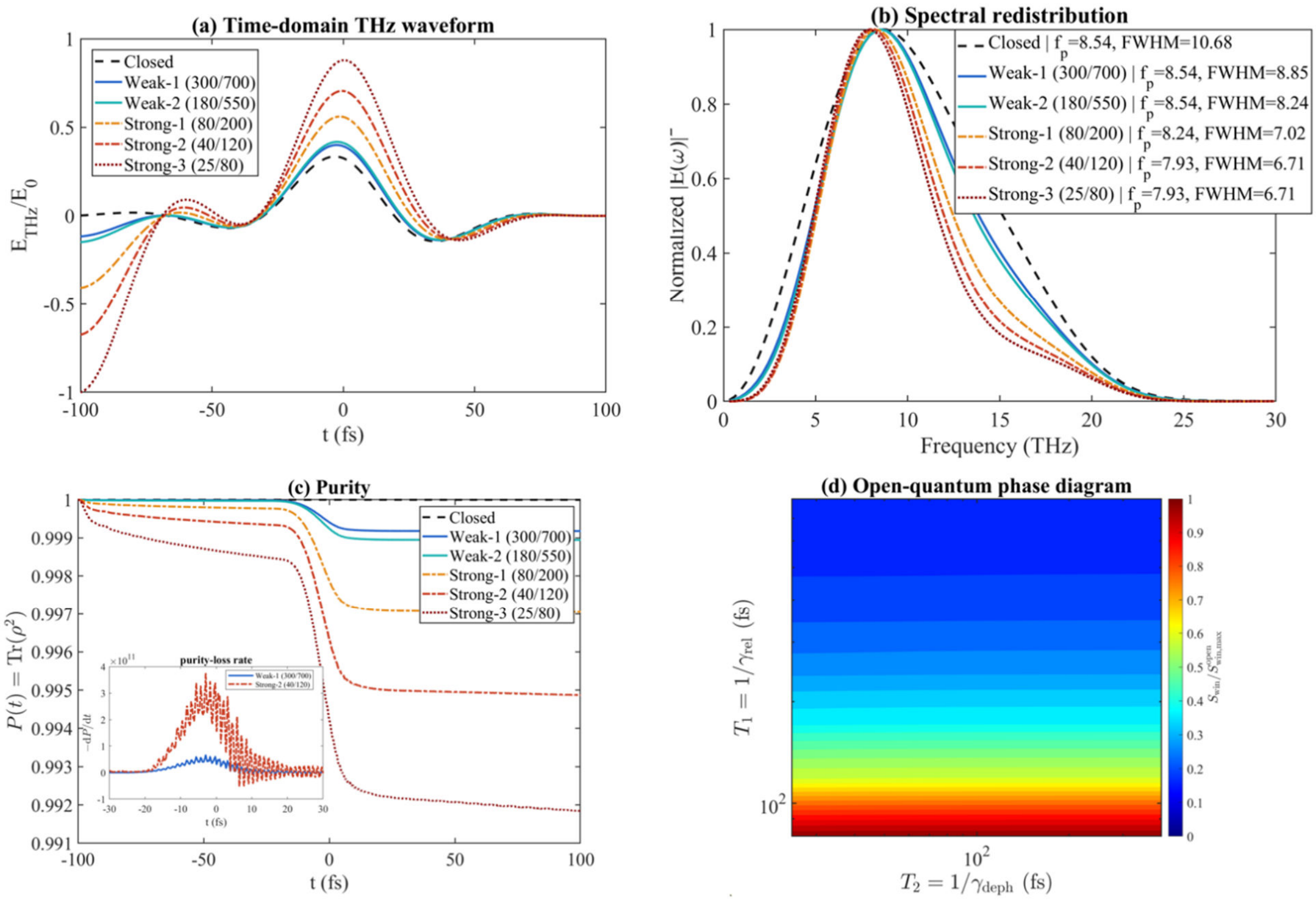

**Figure 6:** Impact of open-quantum relaxation and dephasing on coherent THz emission. (a) Normalized time-domain THz waveforms calculated for different relaxation and dephasing conditions, ranging from the closed-system limit to increasing system–bath coupling. (b) Corresponding normalized THz spectra, illustrating the redistribution of spectral weight, peak frequency, and emission bandwidth induced by open-quantum dynamics. (c) Temporal evolution of the density-matrix purity, $P(t) = \mathrm{Tr}(\rho^2)$, together with the instantaneous purity-loss rate (inset), illustrating the progressive loss of quantum coherence during optical excitation. (d) Open-quantum phase diagram of the normalized integrated THz yield as a function of the longitudinal relaxation time $T_1$ and coherence lifetime $T_2$, identifying the relaxation–dephasing parameter space governing the efficiency of coherent THz generation.

The influence of electronic-band engineering on ultrafast THz generation is illustrated in Fig. 7 through the calculated electronic structures, the corresponding time-domain THz waveforms, and a material–field design map that correlates the normalized THz yield with the effective optical band gap and excitation field strength. The effective electronic band structure provides a direct mechanism for controlling ultrafast THz emission by modifying the microscopic carrier dynamics established during optical excitation. Within the present density-matrix framework, scaling the electronic bands changes the effective optical band gap while simultaneously altering the energetic accessibility of interband transitions, thereby reshaping the transient photocurrent responsible for THz radiation. The calculated band structures demonstrate a systematic increase in the effective optical band gap with increasing band-scaling parameter while preserving the qualitative band ordering and dispersion of the reference model. This scaling provides a controlled way to modify the electronic energy scale and effective optical gap while retaining the qualitative band ordering and dispersion of the reference model. The resulting THz response therefore reflects the combined influence of the effective gap, interband transition energetics, and associated electronic dynamics. The emitted THz waveforms exhibit a distinctly non-monotonic dependence on the engineered electronic structure. The strongest emission is obtained for the reference electronic structure rather than for either the narrowest- or widest-gap cases. This behavior indicates that efficient THz generation is determined by a balance between interband excitation efficiency and coherent carrier transport rather than by maximizing or minimizing the band gap alone. Band-gap engineering therefore modifies the microscopic evolution of the transient photocurrent instead of acting as a simple energetic tuning parameter. The material–field design map identifies an intermediate band-gap regime associated with enhanced normalized THz yield under sufficiently strong optical excitation. Outside this region, both narrow-gap and wide-gap electronic structures produce weaker emission despite identical excitation conditions.

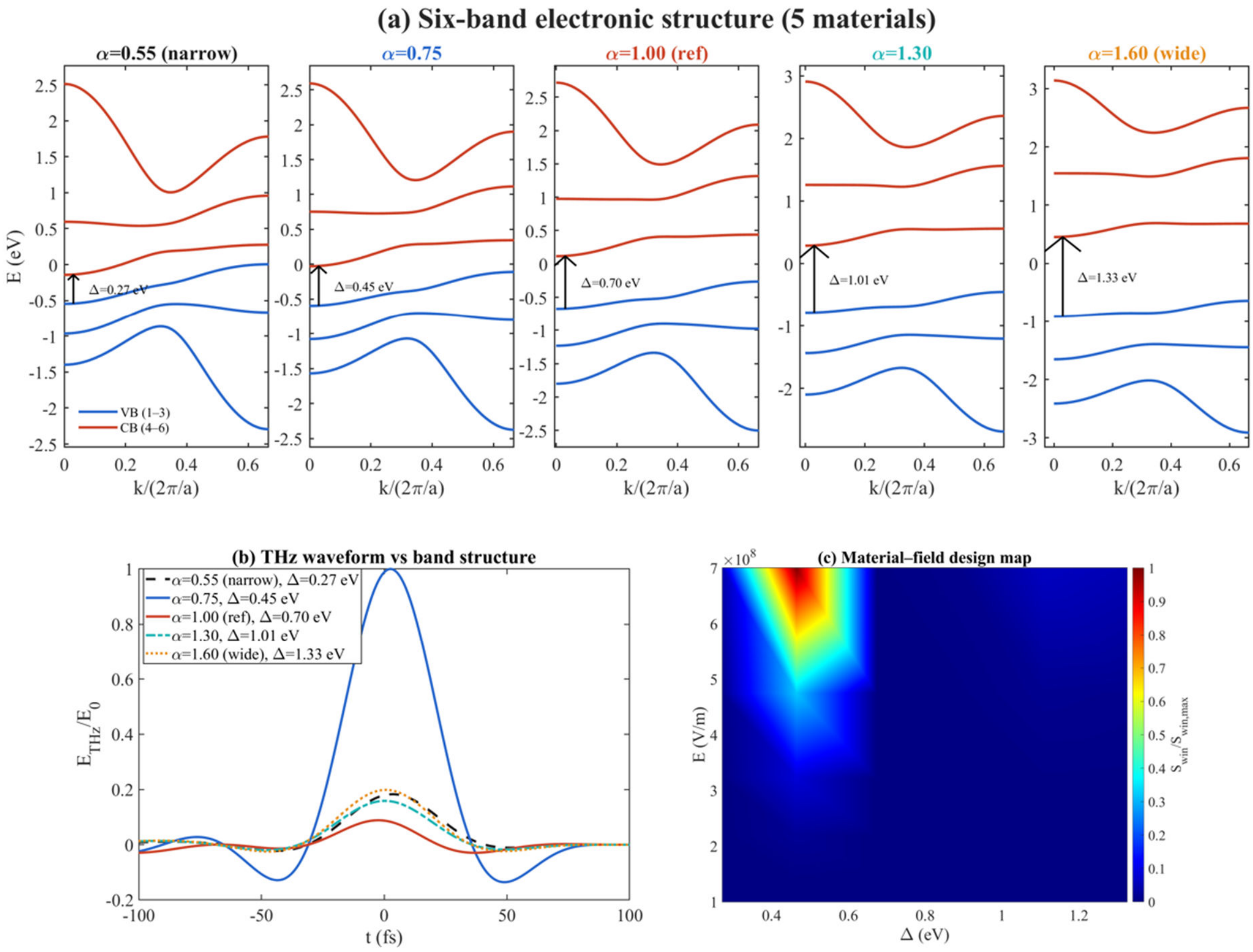

**Figure 7:** Electronic-structure engineering of coherent THz emission through band-gap modulation. (a) Effective six-band electronic structures calculated for different band-scaling parameters ($\alpha$), illustrating the systematic tuning of the effective band gap while preserving the qualitative band ordering and dispersion of the reference model. (b) Normalized THz waveforms generated under identical three-color excitation for the corresponding electronic structures, demonstrating the strong dependence of the emitted field on the engineered band structure. (c) Material–field design map of the normalized integrated THz energy as a function of effective band gap and excitation field strength. The symbols denote the representative electronic structures shown in panel (a), identifying the favorable band-gap regime for efficient coherent THz generation.

The dependence of the favorable band-gap regime on the driving field demonstrates that efficient THz generation is governed by the interplay between the electronic structure and optical excitation, providing a quantitative framework for designing MXene-based THz emitters through electronic-band engineering. The electronic structure of the MXene governs both the efficiency and the spectral characteristics of ultrafast THz emission (Fig. 8). Within the present density-matrix framework, differences among the effective six-band electronic structures reshape the optically accessible interband pathways and thereby control the evolution of nonequilibrium carrier populations, quantum coherences, and the transient photocurrent responsible for THz radiation. The calculated band structures exhibit clear material-to-material variations in effective optical gap and band dispersion (Fig. 8a). These differences modify the strength and energetic accessibility of interband transitions under identical multi-color excitation, indicating that the subsequent carrier dynamics are not determined by the band-gap magnitude alone. Despite identical optical driving, the time-domain THz waveforms remain strongly material dependent (Fig. 8b). Both the emission amplitude and temporal profile vary systematically with composition and surface termination without exhibiting a simple monotonic dependence on the effective optical gap. These material-dependent differences arise from variations in the effective electronic structure and interband transition strengths associated with the different compositions and surface terminations shown in Fig. 8. The radiated THz field is therefore determined by how the electronic structure converts interband excitation into a coherent transient photocurrent rather than by energetic considerations alone. The corresponding spectra show that the same electronic-structure variations also redistribute the spectral weight and emission bandwidth (Fig. 8c). The enhanced high-frequency spectral content observed for selected MXenes is consistent with faster temporal evolution of the microscopic photocurrent supported by their electronic structure. Electronic-structure engineering through surface termination and metal composition provides a direct route for tailoring both the efficiency and the spectral characteristics of coherent THz emission by controlling the nonequilibrium carrier dynamics responsible for transient photocurrent formation.

The microscopic origin of ultrafast THz emission under open-quantum conditions is clarified through the coupled time–frequency evolution of the emitted field and the coherent photocurrent. Within the Redfield density-matrix framework, environmental interactions modify both the temporal coherence of the electronic system and the relaxation of nonequilibrium carrier populations, thereby reshaping the transient current responsible for THz radiation. The time–frequency maps of the emitted THz field in Fig. 9 demonstrate that environmental coupling progressively suppresses the high-frequency components while preserving the overall temporal localization of the emission. The time–frequency maps are normalized independently for each relaxation condition to facilitate comparison of spectral redistribution, rather than absolute emission amplitude. Rather than producing an abrupt spectral transition, relaxation continuously redistributes spectral weight toward lower frequencies, indicating that rapid current oscillations are damped more efficiently than slower collective carrier motion. The evolution of the instantaneous dominant frequency further shows that the characteristic emission frequency remains confined to a narrow temporal interval centered on the optical excitation, while its temporal trajectory is modified by relaxation and dephasing. This behavior indicates that environmental coupling alters the dynamics of frequency formation without changing the fundamental ultrafast timescale of THz generation. The coherent photocurrent exhibits a corresponding evolution in both the time and time–frequency domains. The consistency between the current maps and the emitted-field maps demonstrates the correspondence between the microscopic current dynamics and their radiated-field representation within the present framework. Relaxation suppresses the rapidly varying components of the current, whereas finite coherence allows the dominant emission channel to remain localized during the optical interaction. The direct correspondence between the time–frequency evolution of the coherent photocurrent and the emitted THz field establishes the transient photocurrent as the microscopic quantity governing both the temporal localization and spectral distribution of THz radiation. Open-system dynamics therefore regulate THz emission primarily through the relaxation of nonequilibrium carrier motion and the preservation of coherent current during the ultrafast excitation process.

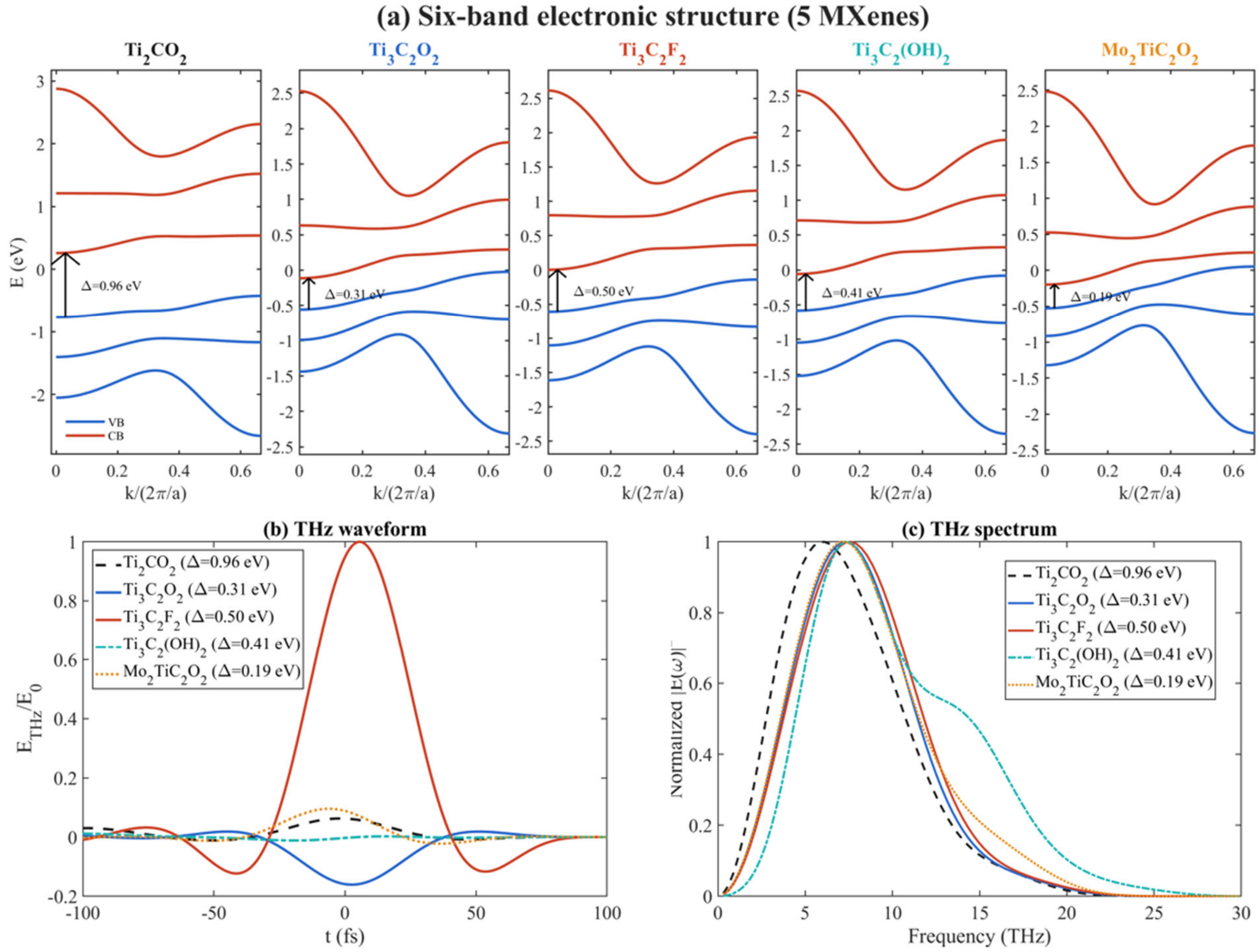

**Figure 8:** Influence of intrinsic MXene electronic structure on coherent THz emission. (a) Effective six-band electronic structures of five representative MXenes, illustrating the variation in band gap and band dispersion across different surface terminations. (b) Normalized THz waveforms generated under identical three-color excitation, demonstrating the pronounced dependence of the emitted field on the intrinsic electronic structure of each MXene. (c) Corresponding normalized THz spectra, revealing material-dependent spectral signatures arising from differences in electronic structure and interband transition strengths.

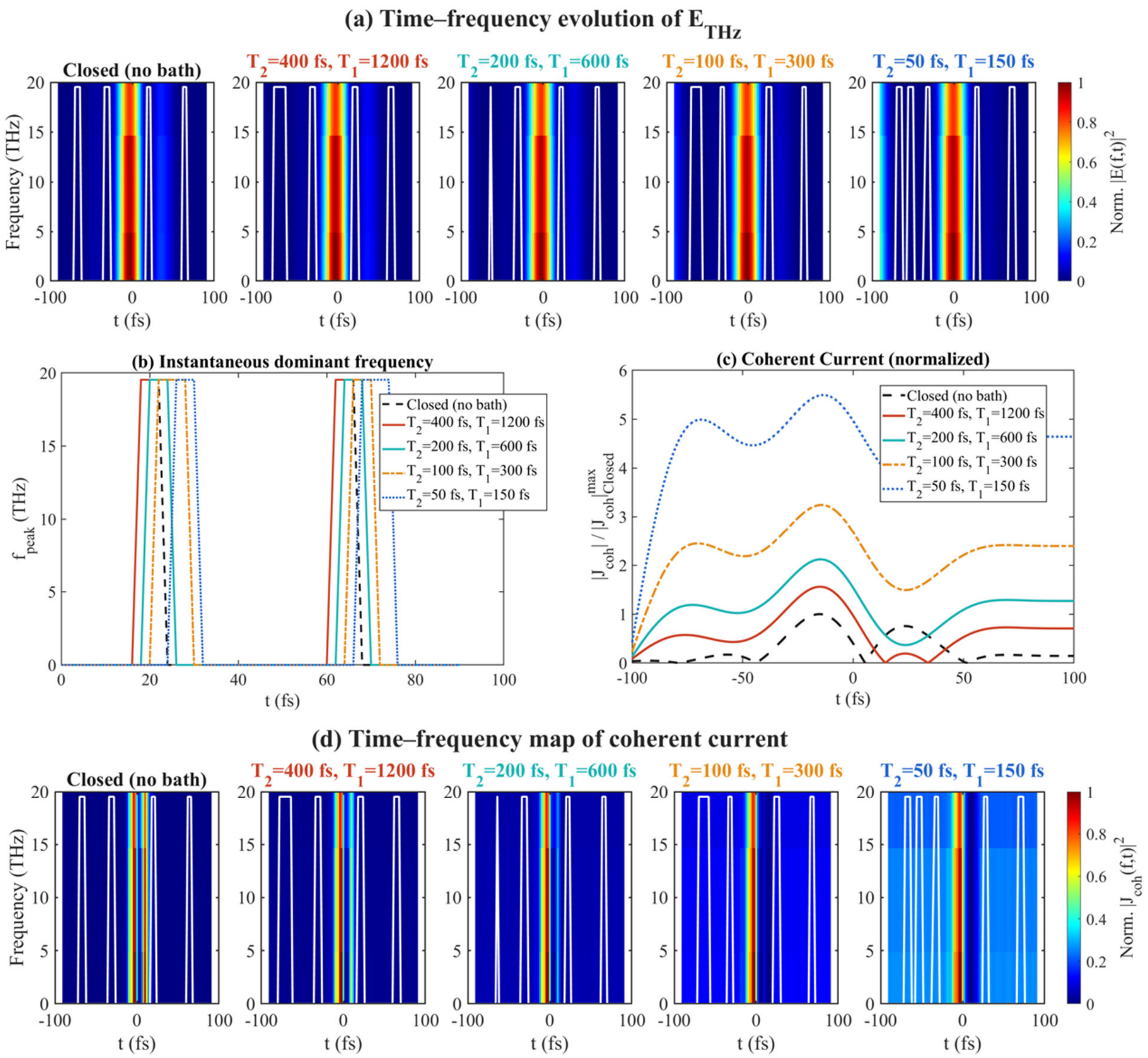

**Figure 9:** Time–frequency evolution of coherent THz emission under open-quantum dynamics. (a) Time–frequency spectrograms of the emitted THz field for different relaxation ($T_1$) and dephasing ($T_2$) times, including the closed-system limit. (b) Instantaneous dominant THz frequency, tracked from the spectral maximum, illustrating the influence of system–bath coupling on the spectral evolution. (c) Normalized coherent photocurrent dynamics for different relaxation and dephasing conditions. (d) Time–frequency spectrograms of the coherent photocurrent, demonstrating the direct correspondence between microscopic current dynamics and the emitted THz radiation. Together, these results show that environmental relaxation redistributes the temporal and spectral evolution of coherent THz emission rather than merely reducing its amplitude.

The microscopic origin of coherent THz emission is analyzed by separating the transient photocurrent into population and coherence-derived components, reconstructing their corresponding radiated fields, evaluating the momentum-dependent interband velocity matrix elements, and mapping the THz transition weight across the electronic band structure (Fig. 10). Here, the transition-resolved THz weight $P_{ij}$ is defined from the time-integrated squared photocurrent associated with each interband channel as $P_{ij} = \int|J_{ij}(t)|^2dt \,/\, \Sigma_{ij}\int|J_{ij}(t)|^2dt$, where $J_{ij}(t)$ denotes the current contribution of the $i \rightarrow j$ transition channel. This normalized quantity provides a direct measure of the relative contribution of each interband pathway to the calculated THz-emission response. This multiscale decomposition establishes a direct connection between density-matrix dynamics and the electronic states responsible for THz radiation. The instantaneous THz source is dominated by the population contribution, while the coherence component remains finite throughout the optical interaction. Population redistribution supplies the largest fraction of the radiating current, whereas quantum coherence reshapes the temporal evolution of the photocurrent and introduces ultrafast features that are absent in the population contribution alone. The reconstructed THz fields reveal that the total emitted waveform cannot be reproduced solely from population dynamics. The difference between the complete response and the population-derived field originates from interband coherence, demonstrating that coherent polarization directly modifies the emitted THz waveform even though its overall contribution is smaller than that of carrier populations. The momentum-resolved electronic structure identifies the microscopic origin of these contributions. Large interband velocity matrix elements are confined to limited regions of the Brillouin zone, showing that only selected electronic states efficiently convert optical excitation into coherent carrier motion. The distribution of transition strength follows the same momentum dependence, indicating that THz generation efficiency is governed by the simultaneous presence of strong optical coupling and favorable electronic velocity. Mapping the THz transition weight onto the electronic band structure further localizes the dominant emission channels. The largest contributions arise from narrow regions of momentum space where interband velocity and electronic dispersion are particularly favourable, while most electronic states contribute only weakly to the radiated field. This microscopic picture establishes a direct relationship between band structure, coherent carrier dynamics, and THz emission, providing a quantitative framework for engineering electronic structures that enhance ultrafast THz generation.

## IV. Conclusion

A multiscale density-matrix formalism has been developed for coherent THz generation in MXenes by integrating parameterized multiband tight-binding electronic structures, non-secular Redfield quantum dynamics, and THz radiation generated from the transient photocurrent within a unified computational scheme. The approach consistently resolves the nonequilibrium evolution of carrier populations and interband coherences under three-color femtosecond excitation, establishing a direct microscopic connection between ultrafast quantum dynamics and coherent THz emission. The results demonstrate that coherent THz radiation is governed not only by nonequilibrium carrier populations but also by the dynamical evolution of interband quantum coherence. Explicit decomposition of the transient photocurrent shows that, although the population contribution dominates under the conditions examined here, quantum coherence provides a finite contribution that modifies the temporal waveform and spectral characteristics of the emitted THz radiation. Environmental interactions further redistribute the coherence contribution in both time and frequency, rather than simply attenuating the emitted signal, thereby modifying the relative balance between coherent and population currents and reshaping the emitted THz waveform. Systematic analyses of optical waveform parameters, effective electronic band structures, and open-system relaxation identify coordinated physical mechanisms governing broadband THz generation. Beyond the specific MXene compositions investigated here, the methodology provides a portable framework that can be adapted to a broader class of low-dimensional quantum materials with appropriate material-specific calibration. It establishes a portable microscopic design platform for connecting MXene-dependent effective electronic structure, quantum coherence, open-system relaxation, nonlinear carrier transport, and coherent

THz emission, while providing physically grounded guidance for waveform engineering, material selection, and the development of broadband coherent THz sources. More broadly, the present approach offers a versatile computational tool for investigating ultrafast light–matter interactions in quantum materials where electronic coherence and environmental coupling must be treated on equal footing.

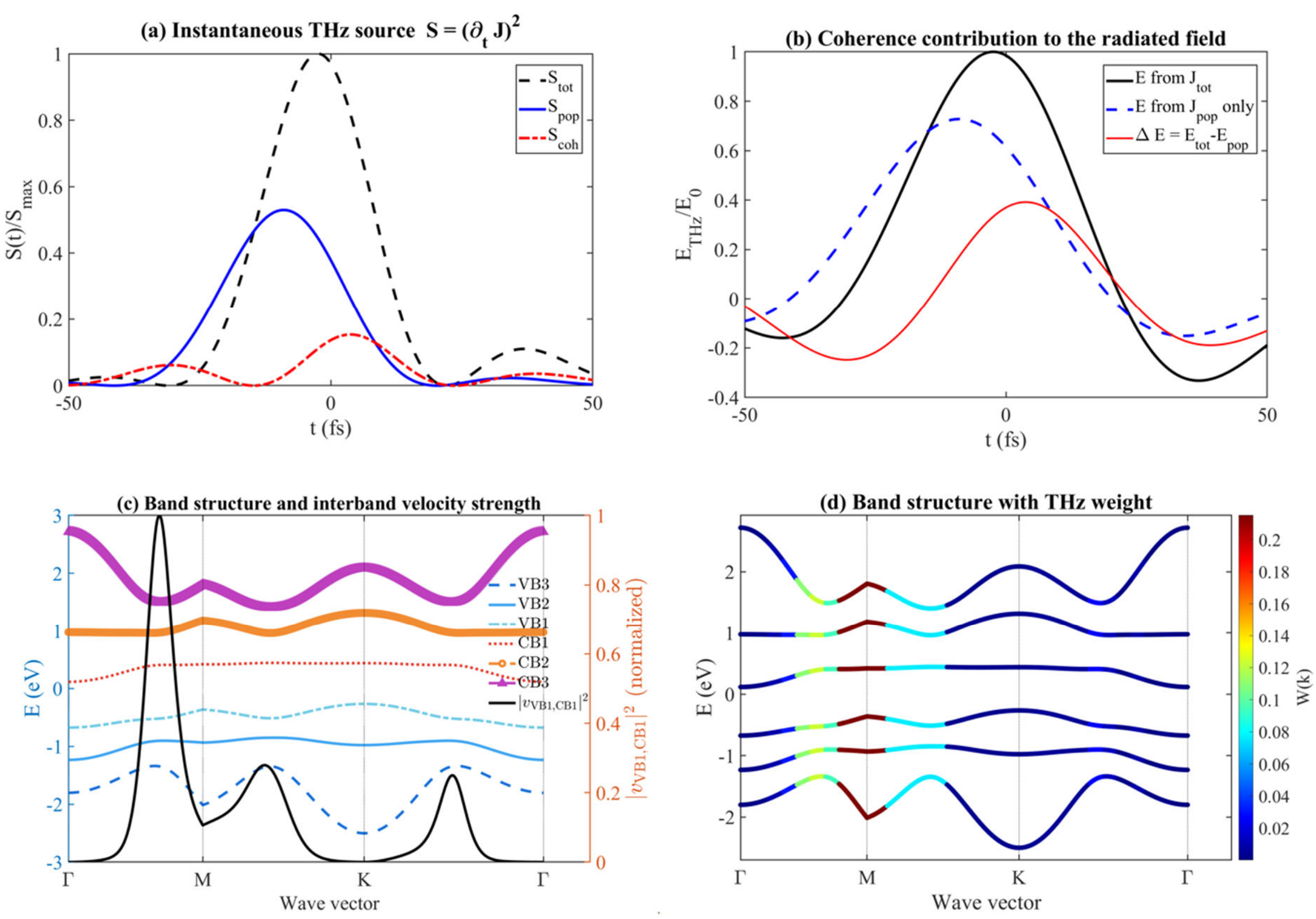

**Figure 10:** Microscopic mechanisms governing coherent THz emission. (a) Instantaneous THz source strength decomposed into total, population-driven, and coherence-driven contributions, illustrating the relative roles of carrier populations and quantum coherences in generating THz radiation. (b) Comparison between the THz field calculated from the total photocurrent and that obtained from the population component alone, with their difference identifying the contribution arising from quantum coherence. (c) Electronic band structure together with the interband velocity matrix element, identifying the optically active transitions that dominate coherent carrier dynamics. (d) Projection of the normalized THz transition weight onto the electronic band structure, revealing the momentum-space regions that contribute most strongly to coherent THz emission.

## Acknowledgements

Some of the authors gratefully acknowledge the financial support provided by the Iran National Science Foundation (INSF) under Grant No. 4030055. Another author gratefully acknowledges the financial support provided by the National Natural Science Foundation of China (NSFC) under Grant No. W2412113.

## Funding

The authors have not disclosed any funding.

## Data Declarations

The author declares no conflicts of interest.

## Data Availability

The data and computational codes supporting the findings of this study are available from the corresponding author upon reasonable request.

## References

[1] H. F. Adeagbo and B. Yang, Non-destructive characterization of drywall moisture content using terahertz time-domain spectroscopy, Sensors 10.3390/s25175576 (2025).

[2] H. Shao, Y. Zhang, Z. Lai, J. Liao, B. Li, W. Wang, P. Xie, B. Chen, C. H. Chan, Y. Meng, and J. C. Ho, Terahertz in-sensor computing utilizing photothermoelectric thin films, Adv. Mater. 10.1002/adma.202505719 (2025).

[3] Z. Liao, Z. Xue, J. Fan, G. Xu, H. Xing, and L. Cong, All-optical hybrid metasurfaces for ultrafast computational spectrometer and single-pixel imaging, Nat. Commun. 10.1038/s41467-025-67605-8 (2025).

[4] C. Sun, J. Li, Q. Liu, X. Chen, J. Leng, L. Du, P. Qi, W. Liu, and L. G. Zhu, Ahigh-speed non-spatial scanning terahertz three-dimensional imaging, Opt. Express 10.1364/OE.575710 (2025).

[5] C. Wang, Y. Tan, Y. Wen, S. Zhao, K. Yu, R. Zhang, J. Sun, and J. Zhou, Multioctave terahertz frequency synthesis via simultaneous excitations of intrinsic and extrinsic nonlinearities in metasurfaces, ACS Nano 10.1021/acsnano.5c15303 (2025).

[6] Z. Hu, S. Sideris, C. McDonnell, T. Ellenbogen, and G. Li, Broadband terahertz holography using nonlinear plasmonic metasurfaces, Nano Lett. 10.1021/acs.nanolett.5c05187 (2025).

[7] Y. Zhang, J. Yang, Y. Zeng, Z. Chen, H. Feng, S. Zhu, K. M. Shum, C. H. Chan, and C. Wang, Monolithic lithium niobate photonic chip for effcient terahertz-optic modulation and terahertz generation, Nat. Commun. 10.1038/s41467-025-65293-y (2025).

[8] B. Han and P. Samorì, Engineering the interfacing of molecules with 2D transition metal dichalcogenides: Enhanced multifunctional electronics, Acc. Chem. Res. 10.1021/acs.accounts.4c00338 (2024).

[9] A. V. Emelianov, M. Pettersson, and I. I. Bobrinetskiy, Ultrafast laser processing of 2D materials: Novel routes to advanced devices, Adv. Mater. 10.1002/adma.202402907 (2024).

[10] A. A. Odebowale, A. M. Berhe, D. Somaweera, H. Wang, W. Lei, A. E. Miroshnichenko, and H. T. Hattori, Advances in 2D photodetectors: Materials, mechanisms, and applications, Micromachines 10.3390/mi16070776 (2025).

[11] J. Rieger, A. Ghosh, J. L. Spellberg, C. Raab, A. Mohan, P. P. Joshi, and S. B. King, Imaging and simulation of surface plasmon polaritons on layered 2D mxenes, Sci. Adv. 10.1126/sciadv.ads3689 (2025).

[12] Z. Fang, Y. Lu, L. Ren, H. Lin, R. Sa, and Z. Ma, Tailoring linear and nonlinear optical properties of 2D sc2c mxenes via surface termination modulation, Dalton Trans. 10.1039/d5dt03041b (2026).

[13] Y. Liu, Y. Wang, N. Wu, M. Han, W. Liu, J. Liu, and Z. Zeng, Diverse structural design strategies of mxene-based macrostructure for high-performance electromagnetic interference shielding, Nanomicro Lett. 10.1007/s40820-023-01203-5 (2023).

[14] R. E. Ustad, S. S. Kundale, K. A. Rokade, S. L. Patil, V. D. Chavan, K. D. Kadam, H. S. Patil, S. P. Patil, R. K. Kamat, D. K. Kim, and T. D. Dongale, Recent progress in energy, environment, and electronic applications of mxene nanomaterials, Nanoscale 10.1039/d2nr06162g (2023).

[15] T. Zhao, H. Wan, T. Zhang, and X. Xiao, Mechanism of the terahertz wave-mxene interaction and surface/interface chemistry of mxene for terahertz absorption and shielding, Acc. Chem. Res. 10.1021/acs.accounts.4c00326 (2024).

[16] H. Hu, S. Wang, Y. Zhang, X. Wang, B. Lin, S. Zhang, W. Zhang, and P. Lian, Mxene-based electromagnetic interference shielding materials: A leap from fundamental research to intelligent customization, Small 10.1002/smll.202505417 (2026).

[17] Y. Zhang, K. Li, and H. Zhao, Intense terahertz radiation: generation and application, Front Optoelectron 10.1007/s12200-020-1052-9 (2021).

[18] S. Chen, L. Zeng, J. Li, J. Weng, J. Li, P. Xu, W. Liu, Y. Sun, J. Yang, Y. Qin, and K. Wen, Multiple plasmoninduced transparency based on black phosphorus and graphene for high-sensitivity refractive index sensing, Opt. Express 10.1364/OE.474901 (2022).

[19] A. A. M. Choobini, S. S. Ghaffari-Oskooi, F. M. Aghamir, and M. Shahmansouri, Mechanisms of THz radiation generation in multi-color laser–plasma interactions: a review across diverse media, Eur. Phys. J. Plus 10.1140/epjp/s13360-026-07369-2 (2026).

[20] A. A. M. Choobini and A. Chimeh, Terahertz wave generation in two-dimensional mxenes under femtosecond pulsed laser illumination, Optical and Quantum Electronics 10.1007/s11082-026-08938-6 (2026).

[21] Z. Guan, B. Wang, G. L. Wang, X. X. Zhou, and C. Jin, Analysis of low-frequency THz emission from monolayer graphene irradiated by a long two-color laser pulse, Opt. Express 10.1364/OE.463568 (2022).

[22] Z. Guan, J. You, B. Wang, X. Li, G. L. Wang, X. X. Zhou, and C. Jin, Generation of polarization-controllable low-frequency THz radiations from single-layer graphene using incommensurate two-color laser pulses, Front. Phys. 10.3389/fphy.2023.1217439 (2023).

[23] M. Kanega and M. Sato, Two-color laser control of photocurrent and high harmonics in graphene, Phys. Rev. B 10.1103/7l1v-1lrm (2025).

[24] K. Jana, A. B. B. de Souza, Y. Mi, S. Gholam-Mirzaei, D. H. Ko, S. R. Tripathi, S. Sederberg, J. A. Gupta, and P. B. Corkum, Terahertz generation via all-optical quantum control in two-dimensional and three-dimensional materials, Phys. Rev. B 10.1103/PhysRevB.111.L161405 (2025).

[25] S. Norman, H. Jung, J. Seddon, S. Prescott, C. T. Harris, S. Addamane, I. Brener, and O. Mitrofanov, Carrier-envelope phase control in terahertz pulse generation using InAs ribbon metasurfaces, ACS Photonics 10.1021/acsphotonics.5c00941 (2025).

[26] Q. Xu, Y. Lu, X. Feng, Q. Wang, L. Niu, L. Liu, X. Zhang, J. Han, and W. Zhang, Recent advances on terahertz wave generation and synchronous control, Advanced Optical Materials 10.1002/adom.202501920 (2026).

[27] G. Li, K. Kushnir, Y. Dong, S. Chertopalov, A. M. Rao, V. N. Mochalin, R. Podila, and L. V. Titova, Equilibrium and nonequilibrium free carrier dynamics in 2D Ti3C2Tx MXenes: THz spectroscopy study, 2D Mater. 10.1088/2053-1583/aacb9e (2018).

[28] G. Li, N. Amer, H. A. Hafez, S. Huang, D. Turchinovich, V. N. Mochalin, F. A. Hegmann, and L. V. Titova, Dynamical control over terahertz electromagnetic interference shielding with 2D Ti3C2Ty MXene by ultrafast optical pulses, Nano Lett. 10.1021/acs.nanolett.9b04404 (2020).

[29] Y. Wu, Y. Wang, D. Bao, X. Deng, S. Zhang, L. Yu-Chun, S. Ke, J. Liu, Y. Liu, Z. Wang, P. Ham, A. Hanna, J. Pan, X. Hu, Z. Li, J. Zhou, and C. Wang, Emerging probing perspective of two-dimensional materials physics: terahertz emission spectroscopy, Light Sci. Appl. 10.1038/s41377-024-01486-2 (2024).

[30] X. Jiang, S. Liu, W. Liang, S. Luo, Z. He, Y. Ge, H. Wang, R. Cao, F. Zhang, Q. Wen, J. Li, Q. Bao, D. Fan, and H. Zhang, Broadband nonlinear photonics in few-layer MXene Ti3C2Tx (T = F, O, or OH), Laser Photonics Rev. 10.1002/lpor.201700229 (2017).

[31] Y. I. Jhon, J. H. Lee, and Y. M. Jhon, Surface termination effects on the terahertz-range optical responses of twodimensional mxenes: Density functional theory study, Materials Today Communications 10.1016/j.mtcomm.2022.103917 (2022).

[32] E. Colin-Ulloa, A. Fitzgerald, K. Montazeri, J. Mann, V. Natu, K. Ngo, J. Uzarski, M. W. Barsoum, and L. V. Titova, Ultrafast spectroscopy of plasmons and free carriers in 2D MXenes, Adv. Material 10.1002/adma.202208659 (2022).

[33] Q. Zhang, J. Li, J. Wen, W. Li, X. Chen, Y. Zhang, J. Sun, X. Yan, M. Hu, G. Wu, K. Yuan, H. Guo, and X. Yang, Simultaneous capturing phonon and electron dynamics in MXenes, Nature Communications 10.1038/s41467-022-35605-7 (2022).

[34] L. Prechtel, L. Song, D. Schuh, P. Ajayan, W. Wegscheider, and A. W. Holleitner, Time-resolved ultrafast photocurrents and terahertz generation in freely suspended graphene, Nature Communications 10.1038/ncomms1656 (2012).

[35] A. M. Buryakov, A. V. Gorbatova, and D. I. Khusyainov, The generation of THz radiation in layered transition metal dichalcogenides, AIP Conference Proceedings 10.1063/5.0055452 (2021).

[36] D. Yagodkin, O. Kirstein, F. Kuhne, A. Scholz, and K. I. Bolotin, Ultrafast photocurrents in MoSe2 probed by terahertz spectroscopy, 2D Materials 10.1088/2053-1583/abd527 (2021).

[37] Y. Gao, Y. Qin, S. Kaushik, E. J. Philip, Y. P. Liu, Y. L. Su, X. Chen, Z. Li, H. Weng, D. E. Kharzeev, M. K. Liu, and J. Qi, Chiral terahertz wave emission from the Weyl semimetal TaAs, Nature Communications 10.1038/s41467-020-14463-1 (2020).

[38] N. Dhakar, S. Kumar, A. Nivedan, and S. Kumar, An interplay between optical rectification and transient photocurrent effect on THz pulse generation from bulk mos2 layered crystal, J. Phys. D: Appl. Phys. 10.1088/1361-6463/ace4d9 (2023).

[39] Y. C. anf X. Lin, C. Wang, Z. Zhang, C. Wang, L. Geng, P. Suo, J. Du, and G. Ma, Revealing a nonlinear photocurrent in the graphene/mos2 heterostructure via terahertz emission spectroscopy, J. Phys. Chem. Lett. 10.1021/acs.jpclett.5c00125 (2025).

[40] E. Mönch, M. D. Moldavskaya, L. E. Golub, V. V. Bel'kov, J. Wunderlich, D. Weiss, J. V. Gumenjuk-Sichevska, C. Niu, P. D. Ye, and S. D. Ganichev, Terahertz radiation driven nonlinear transport phenomena in twodimensional tellurene, Nano Lett. 10.1021/acs.nanolett.4c05279 (2025).

[41] T. Handa, C. Y. Huang, Y. Li, N. Olsen, D. G. Chica, D. D. Xu, F. Sturm, J. W. McIver, X. Roy, and X. Zhu, Terahertz emission from giant optical rectification in a van der waals material, Nat. Mater. 10.1038/s41563-025-02201-1 (2025).

[42] H. Haug and S. W. Koch, *Quantum Theory of the Optical and Electronic Properties of Semiconductors*, 5th ed. (World Scientific, Singapore, 2009).

[43] A. Rudenko, M. K. Hagen, J. Hader, . S. W. Koch, and J. V. Moloney, Self-consistent Maxwell–Bloch model for highorder harmonic generation in nanostructured semiconductors, Photonics Research 10.1364/PRJ.463258 (2022).

[44] A. Dodin, T. Tscherbul, R. Alicki, A. Vutha, and P. Brumer, Secular versus nonsecular Redfield dynamics and Fano coherences in incoherent excitation: An experimental proposal, Phys. Rev. A 10.1103/PhysRevA.97.013421 (2018).

[45] A. Ishizaki and G. R. Fleming, On the adequacy of the Redfield equation and related approaches to the study of quantum dynamics in electronic energy transfer, J. Chem. Phys. 10.1063/1.3155214 (2009).

[46] A. Sindhu, L. E. H. Rodríguez, and A. A. Kananenka, Which redfield theory is the most accurate for simulating 2D electronic spectra?, J. Chem. Phys. 10.1063/5.0318583 (2026).

[47] A. Ghosal and A. K. Roy, A real-time TDDFT scheme for strong-field interaction in cartesian coordinate grid, Chemical Physics Letters 10.1016/j.cplett.2022.139562 (2022).

[48] T. Moitra, Real-time time-dependent density functional theory for pump-probe spectroscopies, arXiv:2605.27252v1 (2026).

[49] M. Lednev, F. J. García-Vidal, and J. Feist, Lindblad master equation capable of describing hybrid quantum systems in the ultrastrong coupling regime, PHYSICAL REVIEW LETTERS 10.1103/PhysRevLett.132.106902 (2024).

# Supplementary Information:

## Quantum–Hydrodynamic Framework of Coherent Terahertz Emission in MXenes Driven via Engineered Femtosecond Waveforms

Ali Asghar Molavi Choobini[1*] Abbas Chimeh[1,2], Jinhui Zhong[3]

[1]Quantum Matter Lab, Department of Physics, College of Science, University of Tehran, Tehran 14399-55961, Iran,

[2]Nexus for Quantum Coherence and Entanglement in Light-Matter Systems (Qcelms), University of Tehran, P.O. Box 14395-547, Tehran, Iran,

[3]Department of Materials Science and Engineering, Southern University of Science and Technology, Shenzhen 518055, China

*Corresponding author: E-mail address: aa.molavich@ut.ac.ir

## I. SUPPLEMENTARY FIGURES

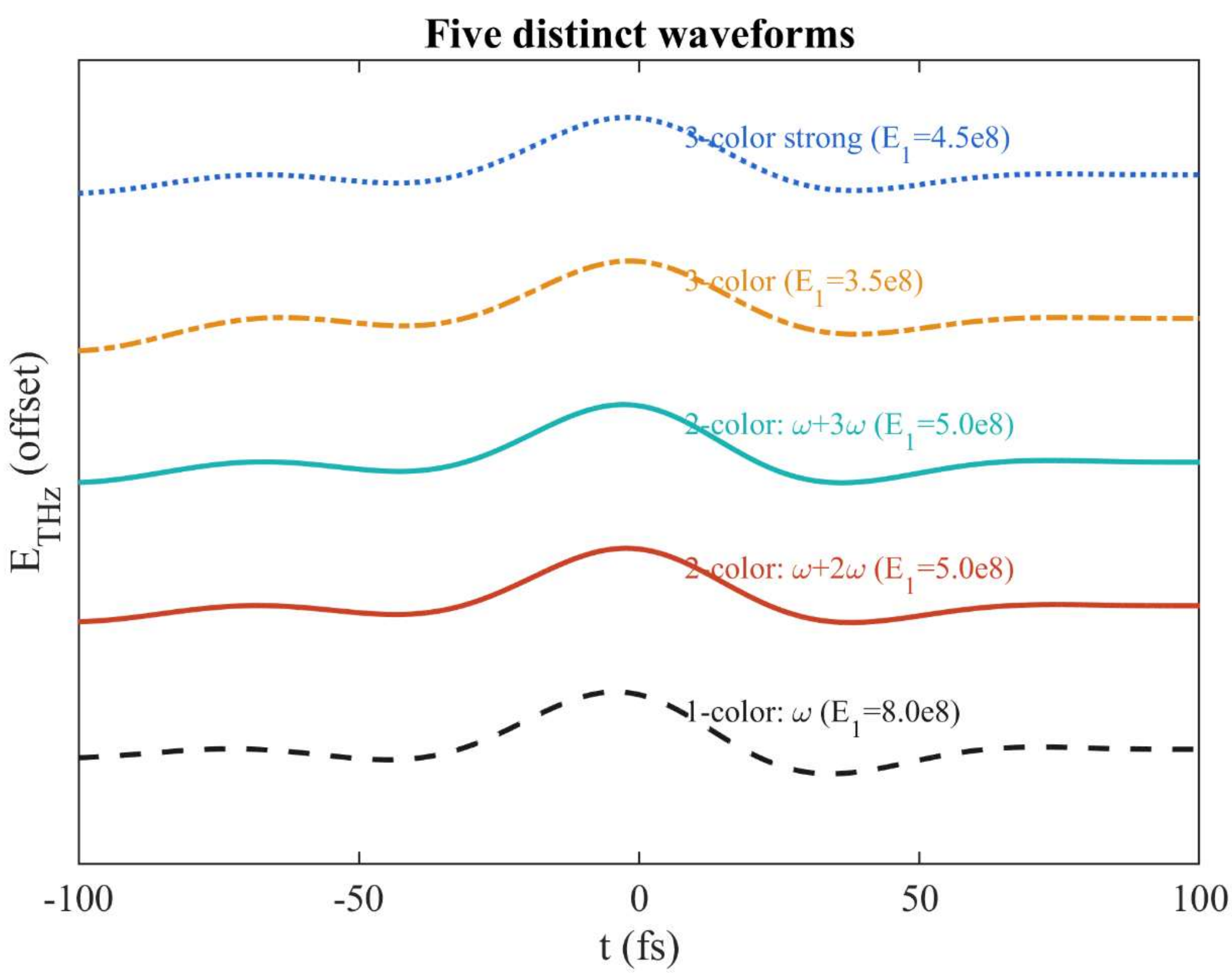

**FIG. 1:** Representative THz waveforms generated under different optical waveform configurations. Comparison of five normalized THz waveforms corresponding to representative single-, two-, and three-color excitation schemes. Vertical offsets are introduced solely for visualization. The results illustrate how different spectral compositions and excitation intensities modify the temporal characteristics of THz emission through controlled nonequilibrium carrier dynamics.

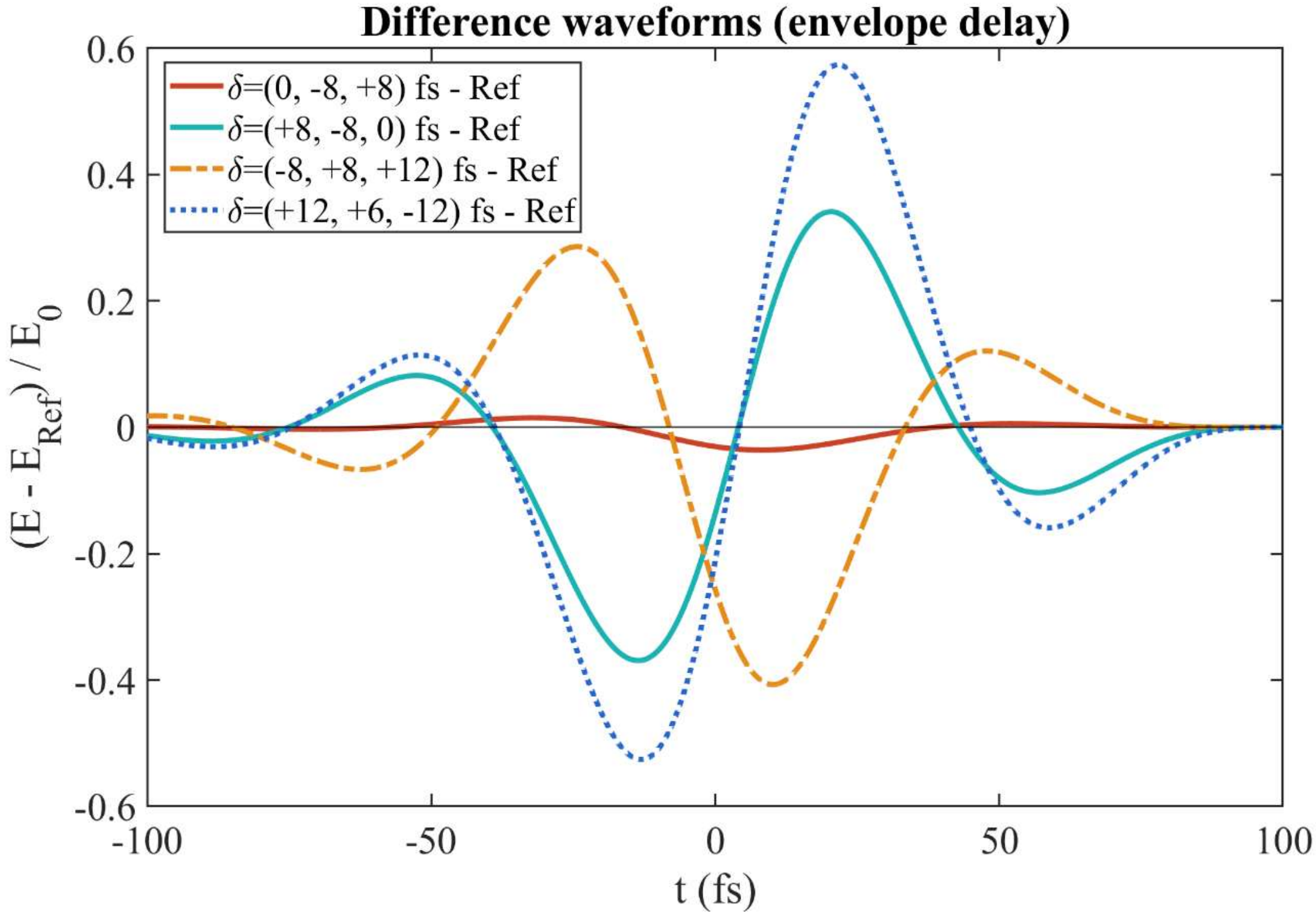

**FIG. 2:** Differential THz waveforms induced by femtosecond envelope-delay engineering. Normalized differential THz waveforms, $(E - E_{\mathrm{Ref}})/E_0$, obtained by subtracting the reference waveform from the responses generated under different envelope-delay configurations. The comparison isolates waveform modifications arising from relative femtosecond temporal delays, revealing how sub-cycle timing redistributes the transient photocurrent and reshapes the emitted THz field without significantly altering the pulse duration.

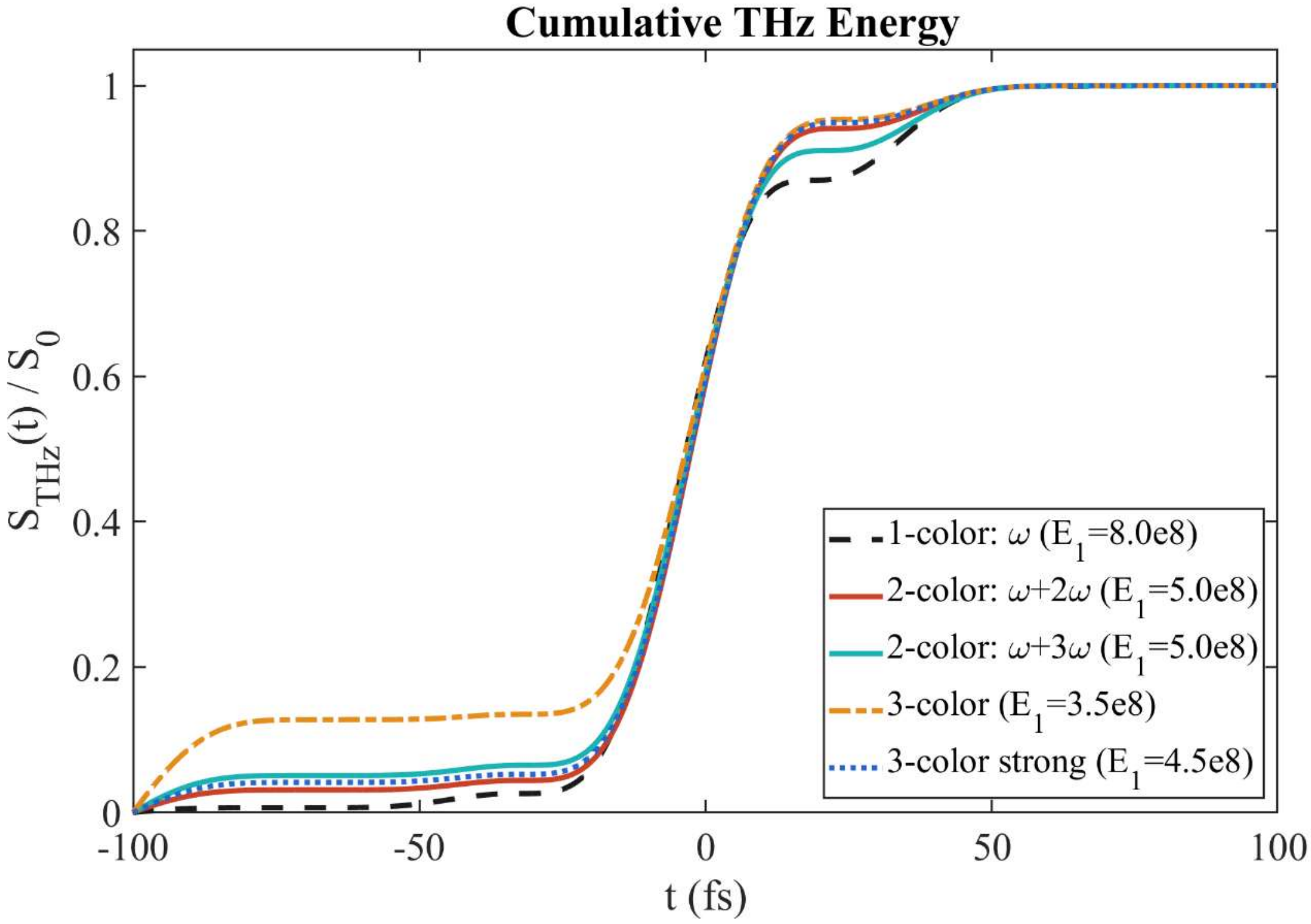

**FIG. 3:** Normalized cumulative THz energy generated under representative single-, two-, and three-color excitation schemes. The cumulative THz energy is obtained by time integration of the emitted THz field and illustrates the temporal accumulation of radiated energy for different waveform-engineering configurations. The curves compare one-color excitation, two-color ($\omega + 2\omega$ and $\omega + 3\omega$), and three-color excitation with different excitation amplitudes.

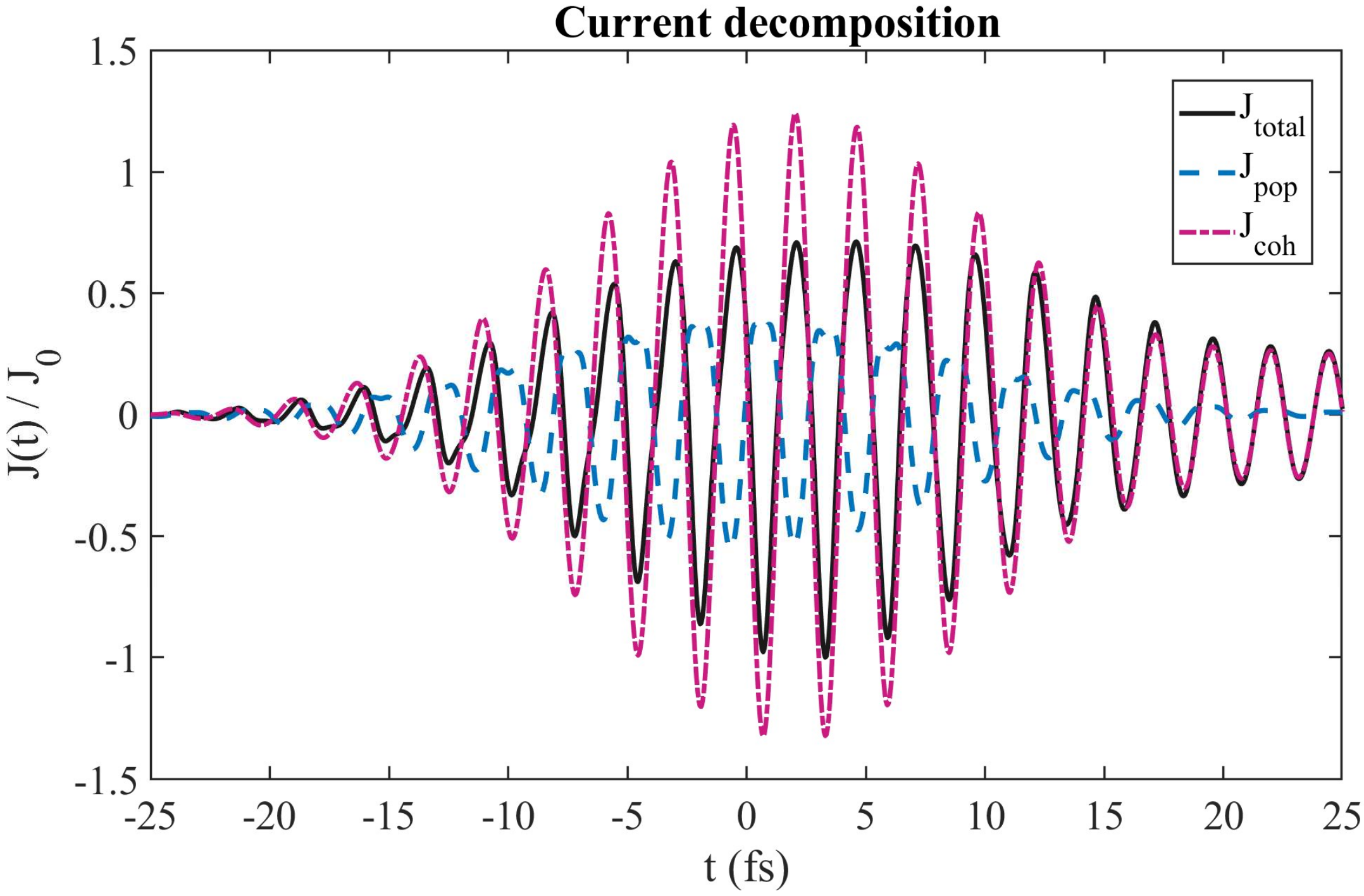

**FIG. 4:** Decomposition of the transient photocurrent into population and coherence contributions. The total transient photocurrent is separated into the population-driven $J_{pop}$ and coherence-driven ($J_{coh}$) components within the Redfield density-matrix framework. The results show that the coherence contribution dominates the peak current during optical excitation, whereas the population contribution evolves more gradually and remains smaller throughout the interaction. Their superposition reproduces the total photocurrent, demonstrating that coherent interband polarization provides the primary microscopic contribution to the ultrafast current responsible for THz emission under the present excitation conditions.

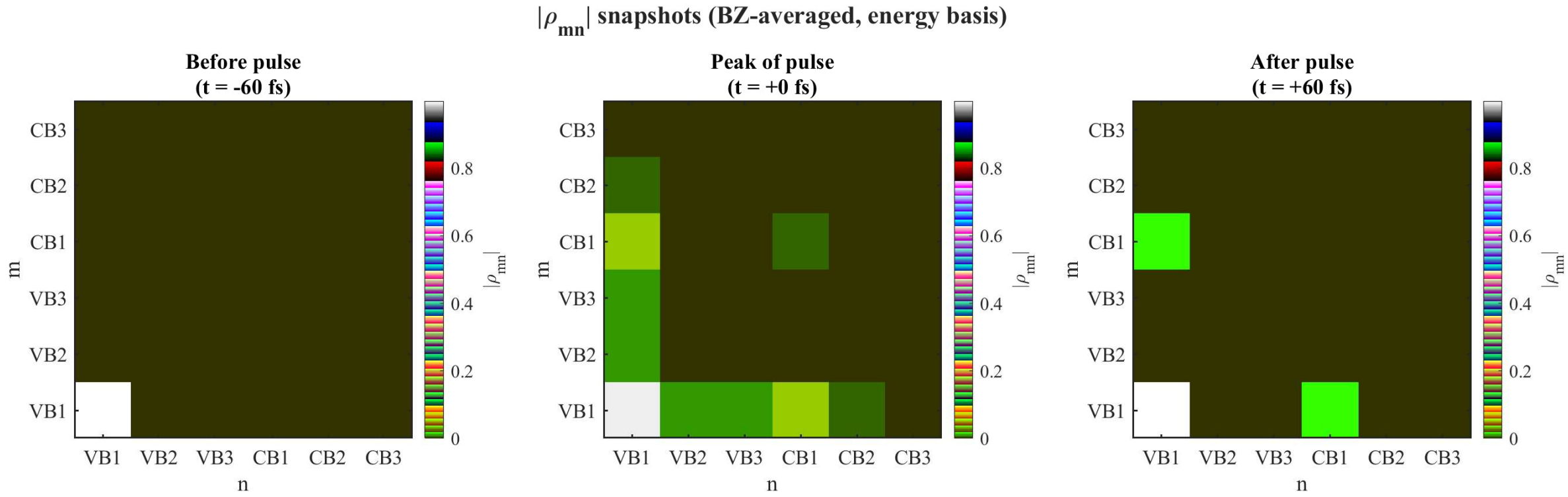

**FIG. 5:** Evolution of the full density matrix during ultrafast optical excitation. Snapshots of the Brillouin-zone-averaged density matrix $|\rho_{mn}|$ in the energy eigenstate basis before the pulse (t = -60 fs), near the pulse maximum (t = 0 fs), and after the pulse (t = +60 fs). Before excitation, the electronic population is localized almost entirely in the lowest valence state. During pulse interaction, population transfer and interband coherences emerge predominantly within the VB1 CB1 manifold, while the remaining matrix elements remain comparatively weak. After the pulse, the populations approach a quasi-stationary distribution with finite residual coherence, illustrating the microscopic quantum evolution underlying the transient photocurrent and the resulting THz emission. This figure complements Fig. 3(c) of the main text by presenting the complete density matrix rather than the enlarged VB1–CB1 subspace, confirming that the dominant quantum dynamics remain localized within the VB1–CB1 manifold throughout the optical excitation.

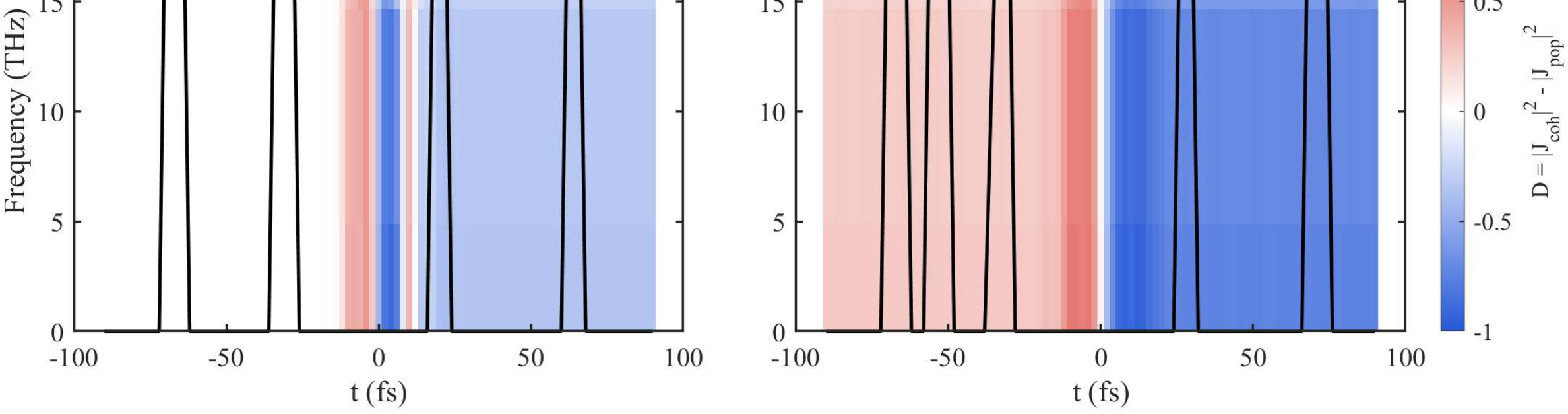

**FIG. 6:** Difference between coherence-driven and population-driven current contributions in the time–frequency domain, $D = |J_{\text{coh}}|^2 - |J_{\text{pop}}|^2$, for (left) the closed-system limit and (right) a representative open-quantum case with finite relaxation and dephasing times. The black curve denotes the driving optical waveform. The comparison illustrates how environmental coupling redistributes the relative importance of coherent and population-mediated current generation during ultrafast THz emission.

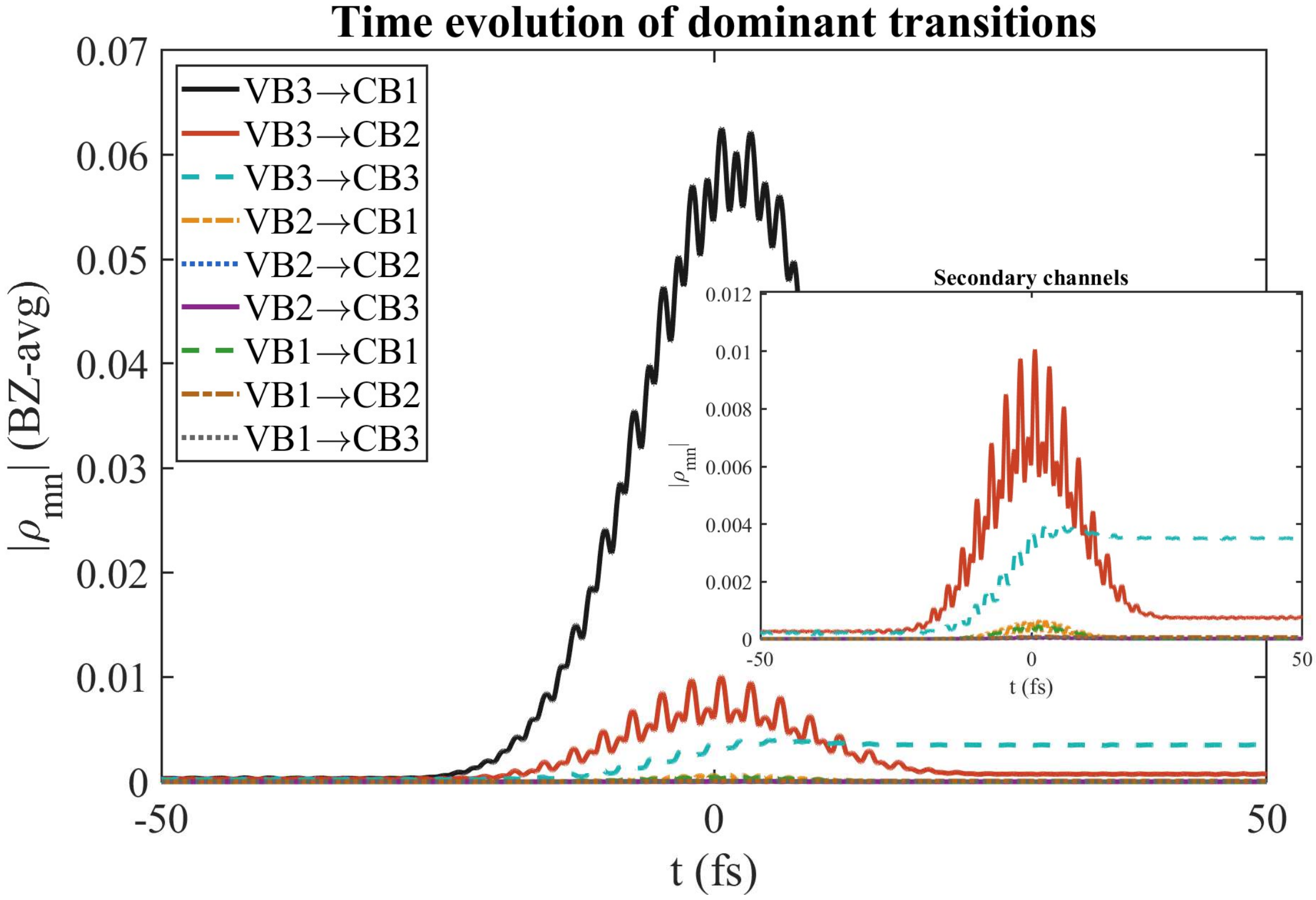

**FIG. 7:** Time evolution of the dominant interband coherence amplitudes obtained from the density-matrix dynamics. The main panel shows the full hierarchy of interband transitions, while the inset enlarges the weaker secondary channels. The results demonstrate that a small number of transitions dominate the coherent carrier dynamics responsible for THz generation, whereas the remaining channels provide only minor corrections.

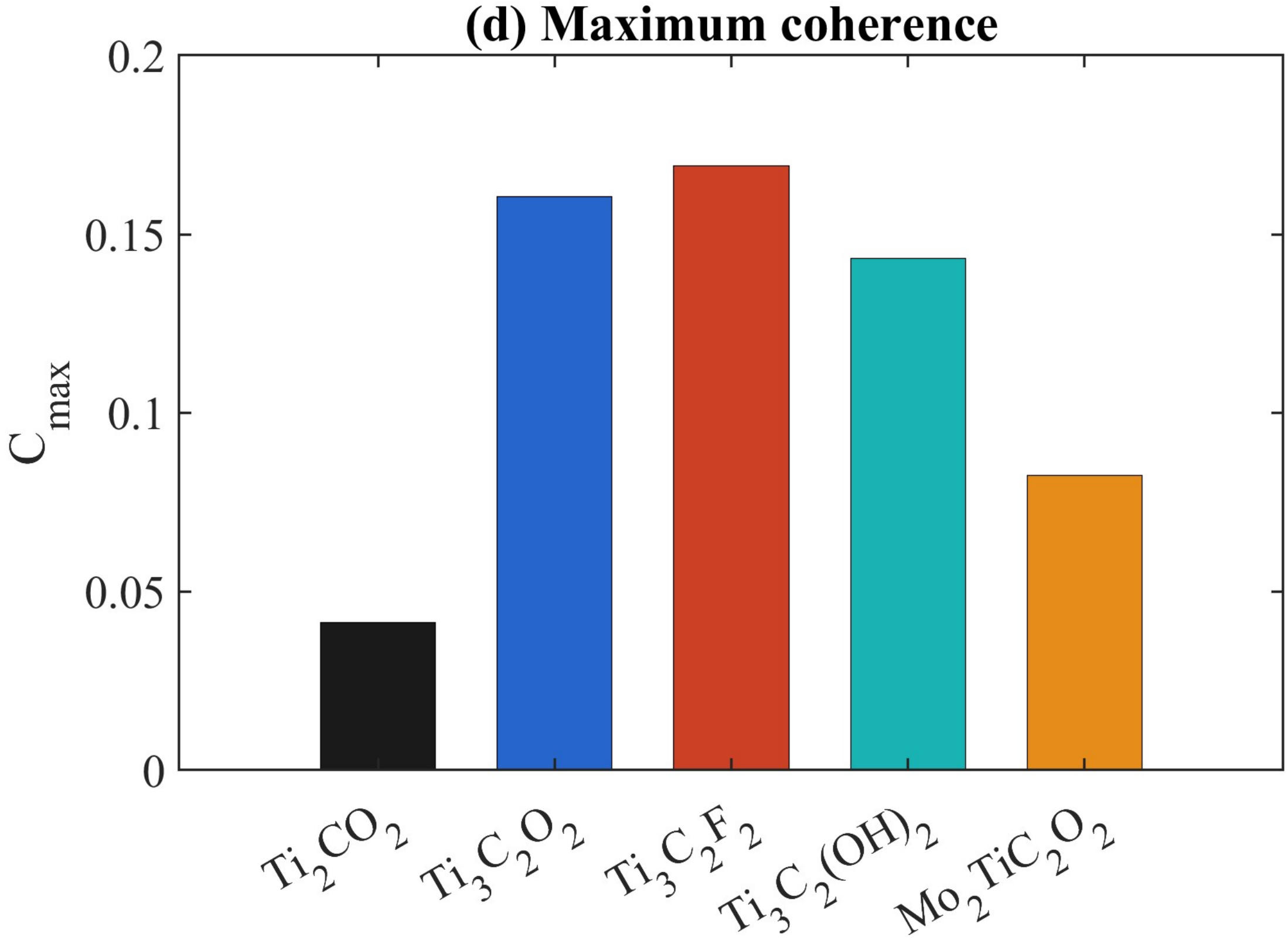

**FIG. 8:** Maximum interband coherence obtained for five representative MXene systems under identical three-color optical excitation. The plotted quantity is the peak value of the dominant coherence amplitude, $C_{\max} = \max_t |\rho\text{VB1,CB1}(t)|$. The comparison demonstrates the strong dependence of coherent carrier polarization on the intrinsic electronic structure of the MXene material.

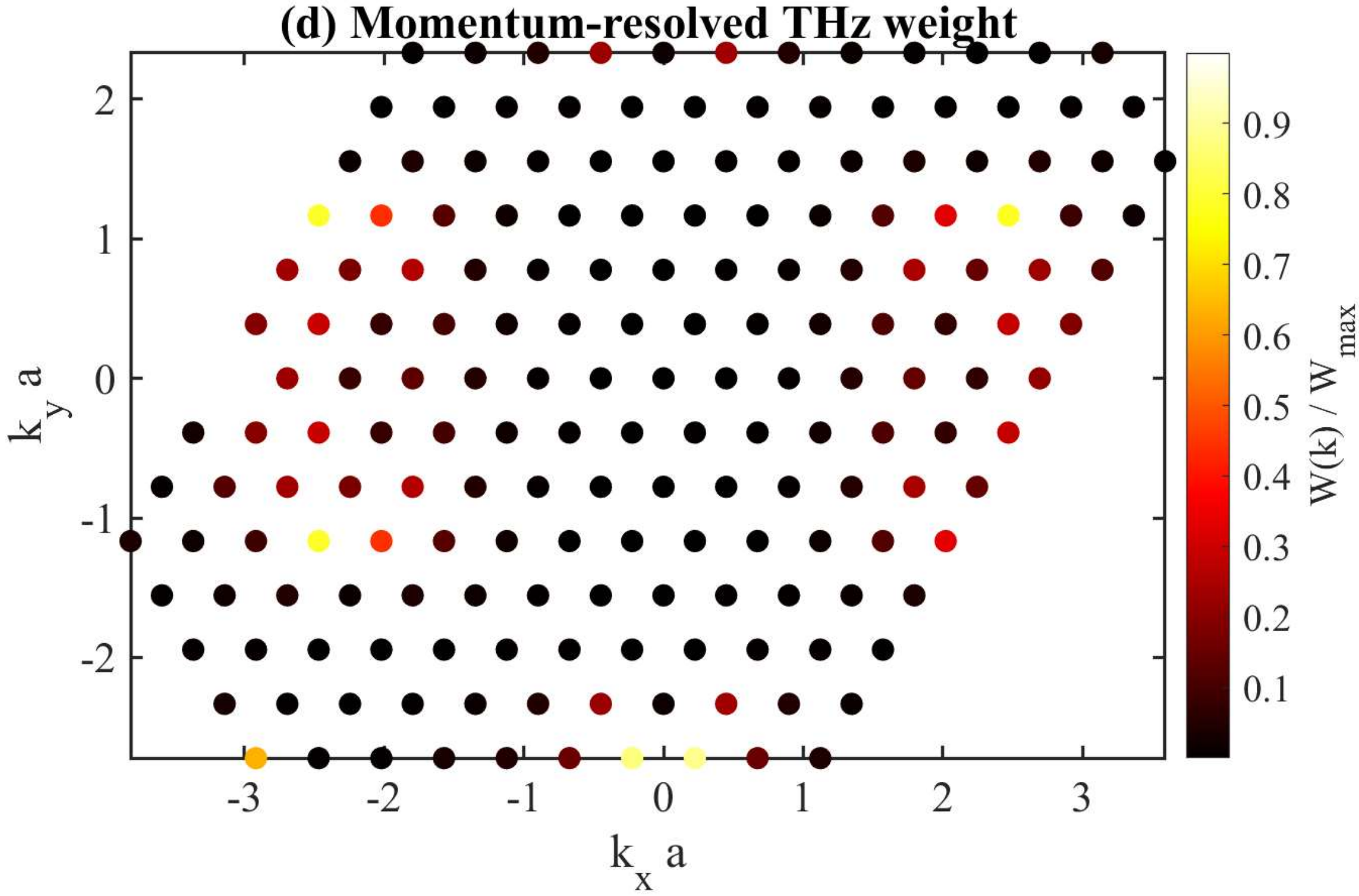

**FIG. 9:** Momentum-resolved distribution of the normalized THz transition weight, $W(\mathbf{k})/W_{\max}$, throughout the Brillouin zone. Bright regions identify the momentum-space locations that contribute most strongly to the emitted THz radiation, revealing that only a limited subset of electronic states dominates the ultrafast emission process.

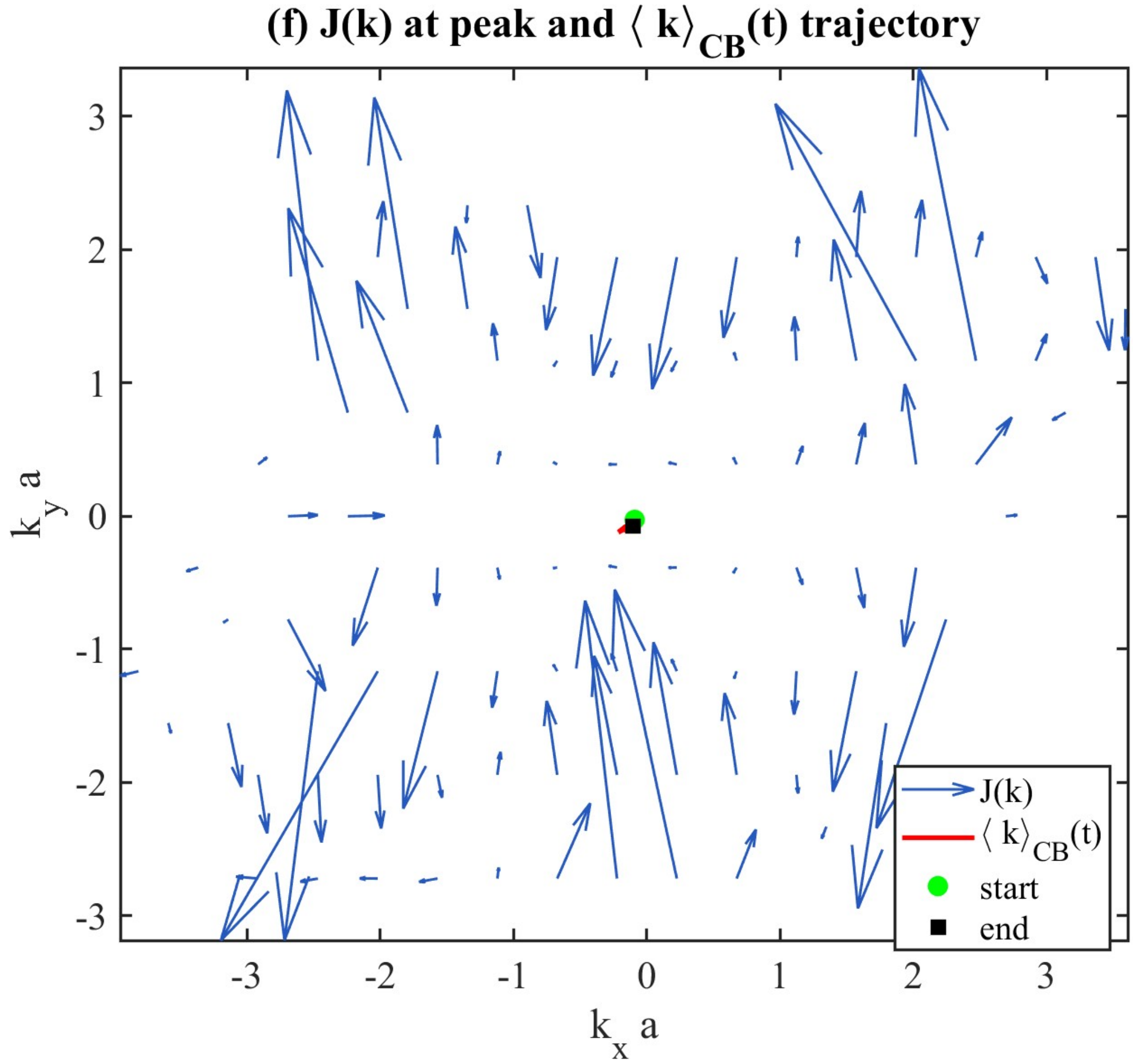

**FIG. 10:** Momentum-space distribution of the microscopic photocurrent at the instant of maximum THz emission. Blue arrows denote the momentum-resolved current density $\mathbf{J}(\mathbf{k})$, while the red trajectory shows the time evolution of the average conduction-band momentum $\langle\mathbf{k}\rangle_{CB}(t)$ during optical excitation. The green and black markers indicate the initial and final states of the trajectory, respectively. The figure illustrates the relation between the transient carrier motion in momentum space and the dominant current responsible for THz radiation.

## II. SUPPLEMENTARY NOTES